\documentclass[12pt]{article}
\usepackage{amsmath,amssymb}
\usepackage{bm}
\usepackage{color}
\usepackage{cite}
\usepackage{mathtools}

\usepackage{multirow}

\usepackage{dsfont}

\DeclareMathOperator{\diag}{diag}

\def\gtap{\ \raisebox{-.4ex}{\rlap{\(\sim\)}} \raisebox{.4ex}{\(>\)}\ }

\usepackage{here}

\newcommand{\id}{\mathds{1}}

\begin{document}

\title{
\begin{flushright}
\ \\*[-80pt]
\begin{minipage}{0.2\linewidth}
\normalsize
\end{minipage}
\end{flushright}
{\Large \bf
$A_4$ Modular Flavour Model of Quark Mass Hierarchies 
		close to the Fixed Point $\tau = i\infty$
\\*[10pt]}}
\author{
		~S. T.~Petcov $^{1,2}$\footnote{Also at:
			Institute of Nuclear Research and Nuclear Energy,
			Bulgarian Academy of Sciences, 1784 Sofia, Bulgaria.}~  
		and~ M. Tanimoto $^{3}$
		\\*[20pt]
{
\begin{minipage}{\linewidth}
$^1${\it \normalsize
INFN/SISSA, Via Bonomea 265, 34136 Trieste, Italy} \\*[5pt]
$^2${\it \normalsize Kavli IPMU (WPI), UTIAS, The University of Tokyo, 
  Kashiwa, Chiba 277-8583, Japan}
\\*[5pt]
$^3${\it \normalsize
Department of Physics, Niigata University, Ikarashi 2, 
Niigata 950-2181, Japan} 
\\*[5pt]
\end{minipage}
}
\\*[40pt]}

\date{
\centerline{\small \bf Abstract}
\begin{minipage}{0.9\linewidth}
\medskip
\medskip
\small
We study the possibility to generate the 
quark mass hierarchies as well as the CKM quark mixing and CP 
violation without fine-tuning in a quark flavour model with modular 
$A_4$ symmetry. The quark mass hierarchies are considered in the vicinity of 
the fixed point $\tau = i\infty$, $\tau$ being the vacuum expectation value  of the modulus. 
We consider first a model in which the up-type and down-type 
quark mass matrices $M_u$ and $M_d$ involve   
modular forms of level 3 and weights 6, 4 and 2 and  
each depends on four constant parameters. Two ratios of these parameters,  
$g_u$ and  $g_d$, can be sources of the CP violation. 
If  $M_u$ and $M_d$ depend on the same $\tau$,
it is possible to reproduce the up-type and down-type 
quark mass hierarchies in the considered model  
for $|g_u|\sim {\cal O}(10)$ with all other constants 
being in magnitude of the same order.
However, reproducing the CP violation in the quark sector is problematic.  
A correct description of the quark mass hierarchies, the quark
mixing and CP violation is possible close to $\tau = i\infty$ 
with all constant being in magnitude of the same order and complex 
$g_u$ and $g_d$, if there are two different moduli $\tau_u$ and $\tau_d$ 
in the up-type and down-type quark sectors.
We also consider the case of $M_u$ and $M_d$ depending on the same $\tau$ and  
involving modular forms of weights 8, 4, 2 and 6, 4, 2, respectively, 
with $M_u$ receiving a tiny SUSY breaking or higher dimensional operator 
contribution.  Both the mass hierarchies of up-type and down-type quarks 
as well and the CKM mixing angles and CP violating phase 
are  reproduced successfully with one complex parameter and 
all parameters being in magnitude of the same order. 
The relatively large value of  ${\rm Im}\,\tau$, needed 
for describing the down-type quark mass hierarchies,  
is crucial for obtaining the correct  up-type quark mass hierarchies.
\end{minipage}
}

\begin{titlepage}
\maketitle
\thispagestyle{empty}
\end{titlepage}

\newpage


%

\newpage
%
\section{Introduction}
\label{Intro}
%
The idea of using modular invariance 
opened up a new promising  direction to challenge the flavour problem
of quarks and leptons \cite{Feruglio:2017spp}.
The main feature of the approach is that the elements of the Yukawa coupling 
and fermion mass matrices in the Lagrangian of the theory are
modular forms of a certain level \(N\) which 
are functions of a single complex scalar field \(\tau\)
-- the modulus -- and have specific transformation properties 
under the action of the modular group.  
The matter fields
(supermultiplets) are assumed to
transform in representations of an inhomogeneous (homogeneous) 
finite modular group \(\Gamma^{(\prime)}_N\), 
while the modular forms furnish
irreducible representations of the same group.
For \(N\leq 5\), the finite modular groups \(\Gamma_N\) 
are isomorphic to the permutation groups 
\( S_3\), \( A_4\), \( S_4\) and \( A_5\)
(see, e.g., \cite{deAdelhartToorop:2011re}), 
while the groups \(\Gamma^\prime_N\) are isomorphic to the double
covers of the indicated permutation groups,
\(S^\prime_3 \equiv S_3\), \(A^\prime_4 \equiv T^\prime\), 
\(S^\prime_4\) and \(A^\prime_5\). These discrete groups 
are widely used in flavour model building.
The theory is assumed to possess the modular symmetry described by the 
finite modular group \(\Gamma^{(\prime)}_N\), which plays the role of a 
flavour symmetry. In the simplest class of such models, 
the vacuum expectation value (VEV) of  modulus \(\tau\) is the only source 
of flavour symmetry breaking, such that no flavons are needed.
%
Another very appealing feature of the proposed framework is that 
the VEV of \(\tau\) can also be the only source of breaking of the 
CP symmetry \cite{Novichkov:2019sqv}.

There is no VEV of $\tau$ which preserves the full modular symmetry.
However, as was  noticed in \cite{Novichkov:2018ovf} 
and further exploited in 
\cite{Novichkov:2018yse,Novichkov:2018nkm,Okada:2020ukr,Okada:2020brs},
there exists three fixed points of the VEV of \(\tau\) 
in the modular group fundamental domain, 
which do not break the modular symmetry completely.
These ``symmetric'' points are
\(\tau_\text{sym} = i\), \(\tau_\text{sym} = 
\omega \equiv \exp(i\,2\pi/ 3)
= -\,1/2 + i\sqrt{3}/2\) (the `left cusp'),
 \(\tau_\text{sym} = i\infty\), and for  
the theories based on the \(\Gamma_N\) 
invariance, preserve, respectively, 
\(\mathbb{Z}^{S}_2\),
\(\mathbb{Z}^{ST}_3\), and 
\(\mathbb{Z}^{T}_N\) residual symmetries. 
In the case of the double cover groups \(\Gamma^\prime_N\),
the \(\mathbb{Z}^{S}_2\) residual symmetry is replaced by 
the \(\mathbb{Z}^{S}_4\) 
and there is an additional \(\mathbb{Z}_2^R\) symmetry
that is unbroken for any value of \(\tau\)
(see~\cite{Novichkov:2020eep} for further details).
When the  flavour symmetry is fully or partially broken, 
the elements of the Yukawa coupling and fermion mass matrices
get fixed and a certain symmetry-constrained flavour structure arises. 

The approach to the flavour problem based on 
modular invariance has been widely explored
so far primarily in the framework 
of supersymmetric (SUSY) theories.
Within the framework of rigid (\(\mathcal{N}=1\)) SUSY,
modular invariance is 
assumed to be a property of 
the superpotential, whose holomorphicity
restricts the number of allowed terms.

Following a bottom-up approach,
phenomenologically viable ``minimal''
lepton flavour models based on modular symmetry,
which do not include flavons,
have been constructed first using the group
\(\Gamma_4 \simeq S_4\)~\cite{Penedo:2018nmg} and
\(\Gamma_3 \simeq A_4\)~\cite{Feruglio:2017spp}
(shown to be describing correctly the available data 
in Ref.\,\cite{Kobayashi:2018scp}).
``Non-minimal'' models with flavons based on
\(\Gamma_2 \simeq S_3\) and \(\Gamma_3 \simeq A_4\)
have been proposed respectively in~\cite{Kobayashi:2018vbk}
and \cite{Criado:2018thu}.

After these studies, the interest in the approach 
grew significantly and a large variety of models has
been constructed and extensively studied.
This includes 
\footnote{A rather complete list of the articles on modular-invariant 
models of lepton and/or quark flavour, which appeared by December of 
2022, can be found in \cite{Petcov:2022fjf}.
We cite here only a representative sample.
}:
%
\begin{itemize} 
%
\item[--] lepton flavour models based on the groups 
\(\Gamma_4 \simeq S_4\)~\cite{Novichkov:2018ovf, Kobayashi:2019mna,Kobayashi:2019xvz,Ding:2019gof},
\(\Gamma_5 \simeq A_5\)~\cite{Novichkov:2018nkm, Ding:2019xna},
\(\Gamma_3 \simeq A_4\)~\cite{Okada:2020brs, Novichkov:2018yse,Ding:2019zxk, Ding:2019gof,Kobayashi:2019gtp,Zhang:2019ngf},
\(\Gamma_2 \simeq S_3\)~\cite{Kobayashi:2019rzp,Okada:2019xqk}
and \(\Gamma_7 \simeq PSL(2,\mathbb{Z}_7)\)~\cite{Ding:2020msi},
\item[--] models of quark flavour~\cite{Okada:2018yrn} and of quark--lepton 
unification~\cite{Kobayashi:2018wkl,deAnda:2018ecu, Okada:2019uoy,Kobayashi:2019rzp,Lu:2019vgm,Okada:2020rjb,Chen:2021zty},
\item[--] models with multiple moduli, considered first phenomenologically
in~\cite{Novichkov:2018ovf, Novichkov:2018yse} and further studied, e.g.,
in~\cite{deMedeirosVarzielas:2019cyj,King:2019vhv,Ding:2020zxw},
\item[--] models in which the formalism of the interplay of modular and 
generalised CP (gCP) symmetries, developed and applied first to the 
lepton flavour problem in~\cite{Novichkov:2019sqv},
is explored~\cite{Kobayashi:2019uyt,Okada:2020brs,Yao:2020qyy,Wang:2021mkw,Ding:2021iqp},
\item[--] models in which current topics (dark matter, leptogenesis, LFV, etc.)  are studied~\cite{ Nomura:2019jxj,Asaka:2019vev,Asaka:2020tmo,Kobayashi:2021jqu,Okada:2021qdf,Tanimoto:2021ehw,Kobayashi:2022jvy,Feruglio:2023uof}.

\end{itemize}
Also the formalism of the double cover finite modular 
groups~\(\Gamma'_N\), to which top-down constructions typically lead
(see, e.g.,~\cite{Nilles:2020nnc,Kikuchi:2020nxn} and references therein),
has been developed and viable flavour models have been constructed
for the cases of \(\Gamma'_3 \simeq T'\)~\cite{Liu:2019khw},
\(\Gamma'_4 \simeq S'_4\)~\cite{Novichkov:2020eep,Liu:2020akv} and 
\(\Gamma'_5 \simeq A'_5\)~\cite{Wang:2020lxk,Yao:2020zml}.
Subsequently these groups have been used for flavour model 
building, e.g., in
 Refs. \cite{Okada:2022kee,Ding:2022nzn,Ding:2022aoe}.
The framework has been further generalised to arbitrary finite modular groups 
(i.e.,~those not described by the series \(\Gamma_N^{(\prime)}\)) 
in Ref.~\cite{Liu:2021gwa}.
It is hoped that the results obtained in the bottom-up 
modular-invariant approach to the lepton and quark flavour
problems will eventually connect with top-down results
(see, e.g.,~\cite{Kobayashi:2021pav,Kobayashi:2018rad, Kobayashi:2018bff, Baur:2019kwi, Kariyazono:2019ehj, Baur:2019iai, Kobayashi:2020hoc, Abe:2020vmv, Ohki:2020bpo, Kobayashi:2020uaj, Nilles:2020kgo,Kikuchi:2020frp, Nilles:2020tdp,  Baur:2020jwc, Ishiguro:2020nuf, Nilles:2020gvu,  Hoshiya:2020hki, Baur:2020yjl,Baur:2021bly,
	Nilles:2021glx,Kikuchi:2021ogn,Li:2021buv,Kobayashi:2021uam,Hoshiya:2022qvr,Kobayashi:2022sov,Kikuchi:2022svo,Kikuchi:2022geu,Kikuchi:2022pkd,Kikuchi:2021yog,Baur:2021mtl,Baur:2022hma}),
based on UV-complete theories.
The problem of modulus stabilisation was also addressed in
\cite{Novichkov:2022wvg,Kobayashi:2019xvz,Kobayashi:2019uyt,Ishiguro:2020tmo,
Knapp-Perez:2023nty}.

The ``minimal'' phenomenologically viable modular-invariant flavour models 
with gCP symmetry constructed so far  
i) of the lepton sector with 
massive Majorana neutrinos (12 observables) contain 
$\geq$ 6 (5) real constants + 1 phase (see, e.g., 
\cite{Novichkov:2018ovf,Novichkov:2019sqv,Ding:2022nzn});
ii) of the quark sector (10 observables) 
contain $\geq$ 8 real parameters and one phase,  
while the models of lepton and quark flavours (22 observables) have 
$\geq$ 14 real parameters and one phase (see, e.g., \cite{Qu:2021jdy}).

 Reproducing correctly  the mass hierarchies of quarks and charged 
leptons is one of the major difficulties in the flavour 
models. 
In almost all phenomenologically 
viable flavour models based on modular invariance constructed 
so far the hierarchy of the charged lepton and quark 
masses is obtained  by severe fine-tuning of some of the 
constant parameters present in the models
\footnote{By fine-tuning we refer to either 
	i) unjustified hierarchies between parameters which are introduced in 
	a model on an equal footing and/or 
	ii) high sensitivity of observables to model parameters.
}.
Perhaps, the only notable exceptions are the non-minimal models 
proposed in Refs. \cite{Criado:2019tzk,King:2020qaj,Kuranaga:2021ujd}, 
in which modular weights are used as Froggatt-Nielsen charges 
\cite{Froggatt:1978nt}, and additional scalar fields of non-zero modular 
weights play the role of flavons.

In Ref. \cite{Novichkov:2021evw} a formalism was developed 
that allows to construct models in which 
the fermion (e.g.~charged-lepton and quark) mass hierarchies 
follow solely from the properties of the modular forms, 
thus avoiding the fine-tuning without the need to introduce extra 
scalar fields. Indeed, in \cite{Novichkov:2021evw} 
the authors have succeeded in constructing a viable lepton flavour model 
in which the charged lepton mass hierarchy is reproduced without 
fine-tuning of the constant parameters present in the charged 
lepton mass matrix.

 The fermion mass matrices are strongly constrained
in each of the three symmetric points  
(i.e., the points of residual symmetries) of the 
VEV of the modulus $\tau$  
in the fundamental domain of the modular group 
discussed above
\cite{Novichkov:2018ovf,Novichkov:2018yse,Novichkov:2018nkm,Okada:2020ukr,Okada:2020brs,Novichkov:2021evw,Feruglio:2021dte,Feruglio:2023bav,Feruglio:2023mii}. 
This fact was exploited in \cite{Novichkov:2021evw} where it was shown 
that fine-tuning
can be avoided in the vicinity 
of the symmetric points $\tau = \omega$ and $\tau = i\infty$
where the charged-lepton and quark 
mass hierarchies can follow from the properties 
of the modular forms present in the corresponding 
fermion mass matrices rather than 
being determined by the values of 
the accompanying constants also present    
in the matrices.
 
Following the  approach in \cite{Novichkov:2021evw} 
quark flavour models based on modular invariance have been proposed in  
\cite{Petcov:2022fjf,Kikuchi:2023cap,Abe:2023ilq,Abe:2023qmr,Kikuchi:2023jap}.
These studies showed, in particular, that 
although it is possible to reproduce 
without fine-tuning the up-type quark or the down-type 
quark mass hierarchies, reproducing  both 
the up-type quark or down-type 
quark masses as well as the 
Cabibbo, Kobayashi, Maskawa (CKM) 
quark mixing and  CP violation 
in the quark sector without fine-tuning 
is a highly challenging problem 
in flavour models with modular symmetry. 

 More specifically, in \cite{Petcov:2022fjf}
we have investigated the possibility 
to describe the  quark mass hierarchies as well as the CKM
mixing and  CP violation  close to symmetric point 
$\tau_\text{sym} = \omega$ in a 
model without flavons with modular $A_4$ symmetry.
In this model the quark doublets furnish a triplet 
irreducible representation of  $A_4$, while the 
right-handed quark fields are (different) singlets of  $A_4$ 
carrying different modular weights.
According to the general results in 
\cite{Novichkov:2021evw}, each of the up-type and down-type quark 
mass hierarchies can be obtained in the model without fine-tuning 
due to the \(\mathbb{Z}^{ST}_3\) residual symmetry. 
The model contains eight constants, only two 
{ ratios} 
of which, $g_u$ and $g_d$, can be a source of the CP violation
in addition to the VEV of the modulus, $\tau = \omega + \epsilon$, 
$(\epsilon)^* \neq \epsilon$, $|\epsilon|\ll 1$.
This is the minimal number of parameters 
ensuring  non-zero values of all quark masses. 
We have shown in \cite{Petcov:2022fjf}, in particular,
that in the case of real (CP-conserving) $g_u$ and $g_d$ and common 
$\tau$ ($\epsilon$) in the up-type  and down-type quark sectors, 
the down-type quark mass hierarchies can be reproduced without fine tuning with 
$|\epsilon| \cong 0.03$ and correspond approximately 
to $1 : |\epsilon| : |\epsilon|^2$,
all other constants being of the same 
order in magnitude. The  up-type quark mass hierarchies, 
which are smaller by a factor $\sim 10$ than the down-type ones, 
can be obtained with the same  $|\epsilon| \cong 0.03$ 
but allowing $|g_u|\sim {\cal O}(10)$ and correspond to  
$1 : |\epsilon|/|g_u| : |\epsilon|^2/|g_u|^2$.
In this setting the correct description 
of the CP violation in the quark sector
represents a severe problem  since it arises 
as a higher order correction in $\epsilon$ 
with $|\epsilon|\ll 1$. This problem can be 
alleviated in the case of 
complex (CP-violating) $g_u$ and  $g_d$ with $|g_u|\sim {\cal O}(10)$, 
but the model failed to reproduce the value of the $V_{cb}$ 
element of the CKM matrix. 
A correct non-fine-tuned description of the quark mass 
hierarchies, the quark mixing  and  CP violation 
is possible with all constants being of the same 
order in magnitude and complex $g_u$ and $g_d$, if one allows 
different values of $\epsilon$ in the up-type  and
down-type quark sectors,  or in a non-minimal 
modification of the considered model which involves 
one modulus and five constants in each 
of the two quark sectors  
one of which is CP violating. 
 
 The results obtained in \cite{Petcov:2022fjf}
encourage us to search for alternative
modular invariant models describing correctly 
without sever fine-tuning the quark mass hierarchies, 
the quark mixing and CP violation.

In this work we study the quark mass hierarchies, the CKM mixing angles and 
the CP violating (CPV) phase close to the fixed point \(\tau_\text{sym} = i\infty\) 
at which the  \(\mathbb{Z}^{T}_N\) residual symmetry holds.
We investigate models based on modular  $A_4$ symmetry ($N=3$) 
since it is rather simple and the modular forms are restricted  
only to triplets and singlets. At  \(\tau_\text{sym} = i\infty\) 
we have \(\mathbb{Z}_3\) residual symmetry as in the case of 
$\tau_\text{sym} = \omega$. However, the 
description of the quark mass hierarchies in the vicinity of  
\(\tau_\text{sym} = i\infty\)  differs significantly from that in the 
vicinity of $\tau_\text{sym} = \omega$ due, in particular, to the significantly 
larger value of ${\rm Im}\,\tau$.

 After introducing the modular symmetry briefly in 
Section~\ref{sec:Modular}, we present the general framework of 
generating the mass hierarchy close to $\tau=i\infty$ without fine-tuning 
in Section~\ref{sec:theory}.
In Section \ref{sec:Model}, we construct a ``minimal''  $A_4$ modular 
invariant model. We then describe how one can naturally generate 
hierarchical quark mass patterns 
in the vicinity of \(\tau_\text{sym} = i\infty\) and 
investigate the flavour structure of the 
quark mass matrices in the model.
After summarizing the input quark data in Section \ref{sec:inputs}, 
we perform numerical analysis of the 
description of the quark mass hierarchies
and CKM parameters by the ``minimal'' model. 
In Section 6.2 we consider phenomenologically 
the case of existence of two different moduli $\tau_u$ and $\tau_d$ 
in the up-type  and down-type quark sectors
and perform a numerical analysis of this case.
We also present in Section 6.3  an 
alternative model in which the quark 
mass matrices depend on a common modulus $\tau$,  
involve altogether eight constant parameters 
(two of which can break the CP symmetry) 
and the lightest u-quark mass is generated 
either by a tiny SUSY breaking contribution  
or by dimension six operators and depends on 
one additional  real parameter. 
We perform also a numerical analysis of the 
compatibility of the model with the existing data.
We summarize our results  in Section~\ref{sec:Summary}.
In Appendix \ref{Tensor}, the decompositions of the tensor 
products  of the irreducible representations of $A_4$ 
are given. In Appendix \ref{fit}, the measure of goodness of
numerical fitting is presented.
Appendix \ref{Det=0} contains a proof 
that the determinant of the up-type quark mass matrix 
considered in Section 6.3 is zero.

\section{Modular symmetry and its Residual Symmetries}
\label{sec:Modular}
%
%
We start by briefly reviewing the modular invariance approach to flavour. 
In the SUSY framework, one introduces a chiral superfield, 
the modulus~\(\tau\), transforming non-trivially under the modular group 
\(\Gamma \equiv SL(2, \mathbb{Z})\). The group~\(\Gamma\) is generated by the 
matrices
\begin{equation}
\label{eq:STR_def}
S =
\begin{pmatrix}
0 & 1 \\ -1 & 0
\end{pmatrix}
\,, \quad
T =
\begin{pmatrix}
1 & 1 \\ 0 & 1
\end{pmatrix}
\,, \quad
R =
\begin{pmatrix}
-1 & 0 \\ 0 & -1
\end{pmatrix}\,,
\end{equation}
%
%
obeying \(S^2 = R\), \((ST)^3 = R^2 = \id\), and \(RT = TR\).
The elements \(\gamma\) of the modular group act on \(\tau\) 
via fractional linear transformations,
\begin{equation}
\label{eq:tau_mod_trans}
\gamma =
\begin{pmatrix}
a & b \\ c & d
\end{pmatrix}
\in \Gamma : \quad
\tau \to \gamma \tau = \frac{a\tau + b}{c\tau + d} \,,
\end{equation}
%
while matter superfields transform as ``weighted'' 
multiplets~\cite{Feruglio:2017spp,Ferrara:1989bc,Ferrara:1989qb},
\begin{equation}
\label{eq:psi_mod_trans0}
\psi_i \to (c\tau + d)^{-k} \, \rho_{ij}(\gamma) \, \psi_j \,,
\end{equation}
%
where \(k \in \mathbb{Z}\) is the so-called modular weight%
\footnote{While we restrict ourselves to integer \(k\), it is also possible 
for weights to be fractional 
\cite{Dixon:1989fj,Ibanez:1992hc,Olguin-Trejo:2017zav,Nilles:2020nnc}.
}
and \(\rho(\gamma)\) is a unitary representation of~\(\Gamma\).

In using modular symmetry as a flavour symmetry, an integer level 
\(N \geq 2\) is fixed and one assumes that \(\rho(\gamma) = \id\) 
for elements \(\gamma\) of the principal congruence subgroup
\begin{equation}
\label{eq:congr_subgr}
\Gamma(N) \equiv
\left\{
\begin{pmatrix}
a & b \\ c & d
\end{pmatrix}
\in SL(2, \mathbb{Z}), \,
\begin{pmatrix}
a & b \\ c & d
\end{pmatrix}
\equiv
\begin{pmatrix}
1 & 0 \\ 0 & 1
\end{pmatrix}
(\text{mod } N)
\right\}\,.
\end{equation}
%
Hence, \(\rho\) is effectively a representation of the (homogeneous) 
finite modular group 
\(\Gamma_N' \equiv \Gamma \, \big/ \, \Gamma(N) \simeq SL(2, \mathbb{Z}_N)\). 
For \(N\leq 5\), this group admits the presentation
\begin{equation}
\label{eq:hom_fin_mod_group_pres}
\Gamma'_N = \left\langle S, \, T, \, R \mid S^2 = R, \, (ST)^3 = \id, \, R^2 = \id, \, RT = TR, \, T^N = \id \right\rangle\,.
\end{equation}
%

The  modulus~\(\tau\) acquires a VEV which is restricted to the upper 
half-plane and plays the role of a spurion, parameterising the breaking of 
modular invariance. Additional flavon fields are not required, and we do not 
consider them here. Since~\(\tau\) does not transform under the \(R\) 
generator, a \(\mathbb{Z}_2^R\) symmetry is preserved in such scenarios 
\cite{Novichkov:2020eep}.
If also matter fields transform trivially under \(R\), one may identify 
the matrices \(\gamma\) and \(-\gamma\), thereby restricting oneself to the 
inhomogeneous modular group~\(\overline{\Gamma} \equiv PSL(2, \mathbb{Z}) \equiv SL(2, \mathbb{Z}) \, / \, \mathbb{Z}_2^R\). 
In such a case, \(\rho\) is effectively a representation of a smaller 
(inhomogeneous) finite modular group \(\Gamma_N \equiv \Gamma \, \big/ \left\langle \, \Gamma(N) \cup \mathbb{Z}_2^R \, \right\rangle\). 
For \(N\leq 5\), this group admits the presentation
\begin{equation}
\label{eq:inhom_fin_mod_group_pres}
\Gamma_N = \left\langle S, \, T \mid S^2 = \id, \, (ST)^3 = \id, \, T^N = \id \right\rangle \,.
\end{equation}
%
In general, however, \(R\)-odd fields may be present in the theory and 
\(\Gamma\) and \(\Gamma_N'\) are then the relevant symmetry groups.
For $N=2,3,4,5$,  $\Gamma_N$, as is well known, 
is isomorphic to the non-Abelian discrete symmetry groups
$S_3$, $A_4$, $S_4$, $A_5$ and  $\Gamma_N'$ is isomorphic to their 
respective double covers.

Finally, to understand how modular symmetry may constrain the Yukawa couplings 
and mass structures of a model in a predictive way, we turn to the 
Lagrangian -- which for an \(\mathcal{N} = 1\) global supersymmetric theory 
is given by
\begin{equation}
\mathcal{L} = \int \text{d}^2 \theta \, \text{d}^2 \bar{\theta} \, K(\tau,\bar{\tau}, \psi_I, \bar{\psi}_I)
+ \left[ \, \int \text{d}^2 \theta \, W(\tau,\psi_I) + \text{h.c.} \right] \,.
\end{equation}
%
Here \(K\) and \(W\) 
are the K\"ahler potential and the superpotential, respectively. The superpotential \(W\) 
can be expanded in powers of matter superfields \(\psi_I\),
\begin{equation}
\label{eq:W_series}
W(\tau, \psi_I) = \sum \left( \vphantom{\sum} Y_{I_1 \ldots I_n}(\tau) \, \psi_{I_1} \ldots \psi_{I_n} \right)_{\mathbf{1}} \,,
\end{equation}
%
where one has summed over all possible field combinations and 
independent singlets of the finite modular group.
By requiring the invariance of the superpotential under modular 
transformations,
one finds that the field couplings \(Y_{I_1 \ldots I_n}(\tau)\) have to be 
modular forms of level \(N\). These are severely constrained holomorphic 
functions of~\(\tau\), which under modular transformations obey
\begin{equation}
\label{eq:Y_mod_trans}
Y_{I_1 \ldots I_n}(\tau) \,\xrightarrow{\gamma}\, Y_{I_1 \ldots I_n}(\gamma \tau) = (c\tau + d)^{k} \rho_Y(\gamma) \,Y_{I_1 \ldots I_n}(\tau) \,.
\end{equation}
%
Modular forms carry weights \(k = k_{I_1} + \ldots + k_{I_n}\) and furnish 
unitary irreducible representations \(\rho_Y\) of the finite modular group such 
that \(\rho_Y \,\otimes\, \rho_{I_1} \,\otimes \ldots \otimes\, \rho_{I_n} \supset \mathbf{1}\).
Non-trivial modular forms of a given level exist only for  
\(k \in \mathbb{N}\), span finite-dimensional linear spaces
\(\mathcal{M}_{k}(\Gamma(N))\), and can be arranged into multiplets 
of \(\Gamma^{(\prime)}_N\).

The breakdown of modular symmetry is parameterised by the VEV of the modulus 
and there is no value of \(\tau\) which preserves the full symmetry. 
Nevertheless, at certain so-called symmetric points \(\tau = \tau_\text{sym}\) 
the modular group is only partially broken, with the unbroken generators 
giving rise to residual symmetries.
In addition, as we have noticed, the \(R\) generator is unbroken for any 
value of \(\tau\), so that a \(\mathbb{Z}_2^R\) symmetry is always preserved.
There are only three inequivalent symmetric points (in the fundamental domain
of the modular group), namely~
\cite{Novichkov:2018ovf}:
\begin{itemize}
	\item \(\tau_\text{sym} = i \infty\), invariant under \(T\), 
preserving \(\mathbb{Z}_N^T \times \mathbb{Z}_2^R\)\,,
	\item \(\tau_\text{sym} = i\), invariant under \(S\), preserving \(\mathbb{Z}_4^S \) (recall that \(S^2 = R\))\,,
	\item \(\tau_\text{sym} = \omega \equiv \exp (2\pi i / 3)\), `the left cusp', invariant under \(ST\), preserving \(\mathbb{Z}_3^{ST} \times \mathbb{Z}_2^R\)\,.
\end{itemize}


\section{Mass hierarchy without fine-tuning close to $\tau=i\infty$}
\label{sec:theory}

In theories where modular invariance 
is broken only by the VEV of modulus $\tau$, 
the fermion mass matrices (in the limit of unbroken SUSY)
are expressed in terms of modular forms of a given level $N$
 and a limited number 
of couplings in the superpotential. 
The flavour structure of the mass matrices 
is determined by the properties of the respective modular forms 
at the value of the VEV of \(\tau\). It  can be severely constrained 
at the points of residual symmetries \(\tau = \tau_\text{sym}\) 
which typically enforce the presence of multiple zero entries 
in the mass matrices due to zero values of the corresponding modular 
form components. As \(\tau\)  moves away from its symmetric value, 
these entries will generically become non-zero 
but are suppressed and thus a flavour structure arises. 

This approach to fermion (charged lepton and quark) mass hierarchies was 
explored in, e.g., \cite{Feruglio:2021dte,Novichkov:2021evw,Petcov:2022fjf}.
We present below a more detailed discussion of the approach   
following \cite{Novichkov:2021evw}.
If $\epsilon$ parameterises  the deviation of  \(\tau\)  from a
given symmetric point \(\tau_\text{sym}\), 
$|\epsilon| \ll 1$, the degree of suppression 
of the (residual-)symmetry-breaking entries 
is determined by $|\epsilon|^l$, 
where $l > 0$ is an integer. 
The integer constant $l$ is not another 
free parameter of the theory. The values $l$ can take 
depend on the symmetric point, i.e., on the 
residual symmetry. As was shown in \cite{Novichkov:2021evw}, 
i) for $\tau_\text{sym} = i$, $l=0,1$,
ii) for $\tau_\text{sym} = \omega$, $l=0,1,2$, 
and iii) for $\tau_{\rm sym} = i\infty$ 
of interest and  $\Gamma_N$ ($\Gamma^\prime_N$) finite modular group, 
$N\leq 5$, $l$ can take values $l = 0,1,...,N-1$.
Thus, for  $\Gamma_3 \simeq A_4$ and  $\tau_{\rm sym} = i\infty$, 
$l = 0,1,2$. For a specific would-be-zero element of the 
fermion mass matrix the value of $l$ is uniquely 
determined by the residual symmetry  
and by the transformation properties of the fermion fields, 
associated with the entry,  under the 
residual symmetry group  \cite{Novichkov:2021evw}.

Indeed, consider the  modular-invariant bilinear
\begin{equation}
\label{eq:bilinear}
\psi^c_i \, M(\tau)_{ij}\, \psi_j \,,
\end{equation}
%
where the superfields \(\psi\) and \(\psi^c\) transform under the 
modular group as%
\footnote{Note that in the case of a Dirac bilinear \(\psi\) and \(\psi^c\) 
are independent fields, so in general \(k^c \neq k\) and 
\(\rho^c \neq \rho, \rho^{*}\).}
\begin{equation}
\label{eq:psi_mod_trans}
\begin{split}
\psi \,&\xrightarrow{\gamma}\, (c \tau + d)^{-k} \rho(\gamma) \,\psi \,, \\
\psi^c \,&\xrightarrow{\gamma}\, (c \tau + d)^{-k^c} \rho^c(\gamma)\, \psi^c \,,
\end{split}
\end{equation}
%
so that each \(M(\tau)_{ij}\) is a modular form of level \(N\) and 
weight \(K \equiv k+k^c\).
Modular invariance requires \(M(\tau)\) to transform as
\begin{equation}
\label{eq:mass_matrix_mod_trans}
M(\tau)\, \xrightarrow{\gamma}\, M(\gamma \tau) 
= (c \tau + d)^K \rho^c(\gamma)^{*} M(\tau) \rho(\gamma)^{\dagger} \,.
\end{equation}
%
Taking \(\tau\) to be close to the symmetric point, and setting \(\gamma\) 
to the residual symmetry generator, one can use this transformation 
rule to constrain the form of the mass matrix \(M(\tau)\) 
\cite{Novichkov:2021evw}. 

At \(\tau_\text{sym} = i\infty\) of interest we have  
$Z^T_N$ residual symmetry group gnerated by the $T$ generator of 
$\Gamma_{\rm N}$ ($\Gamma^\prime_{\rm N}$).
Consider the $T$-diagonal representation basis for 
group generators in which \(\rho^{(c)}(T) =\diag (\rho^{(c)}_i)\), 
and assume that \(\tau\) is ``close'' to
\(\tau_\text{sym} = i\infty\), i.e., that ${\rm Im}\,\tau$ 
is sufficiently large. 
By setting \(\gamma = T\) in Eq.\,\eqref{eq:mass_matrix_mod_trans}, 
one finds
\begin{equation}
\label{eq:mass_matrix_T_trans}
M_{ij}(T \tau) = \left( \rho^c_i\, \rho_j \right)^{*} M_{ij}(\tau) \,.
\end{equation}
%
It is now convenient to treat the \(M_{ij}\) as functions of 
$\hat q \equiv \exp(i2\pi \tau/N)$, so that 
\begin{equation}
\epsilon  \equiv |\hat q| = e^{-\,2\pi {\rm Im}\tau/N} \, 
\label{eq:q}
\end{equation}
%
parametrises the deviation of \(\tau\) 
from the symmetric point \cite{Novichkov:2021evw}. 
Note that the entries \(M_{ij}(\hat q)\) depend analytically on \(\hat q\) and that 
\(\hat q \xrightarrow{T} \xi \hat q\), with $\xi = \exp(i\,2\pi/N)$.
Thus, in terms of \(\hat q\), 
Eq.\,\eqref{eq:mass_matrix_T_trans} reads:
\begin{equation}
M_{ij}(\xi \hat q) =  (\rho^c_i\, \rho_j)^{*} M_{ij}(\hat q)\,.
\end{equation}
%
Expanding both sides in powers of \(\hat q\), one finds:
\begin{equation}
\label{eq:expansion_ST}
\xi^{n} M_{ij}^{(n)}(0) = 
(\rho^c_i \, \rho_j)^{*} M_{ij}^{(n)}(0)\,,
\end{equation}
%
where \(M_{ij}^{(n)}\) denotes the \(n\)-th derivative of 
\(M_{ij}\) with respect to \(\hat q\).
It follows that \(M^{(n)}_{ij}(0)\) can only be non-zero for values 
of $n$ such that $(\rho^c_i\, \rho_j)^{*} = \xi^n$. In particular,
the entry \(M_{ij} = M^{(0)}_{ij}(0)\)
is only allowed to be non-zero and be \(\mathcal{O}(1)\) 
if {\(\rho^c_i \rho_j = 1\)}. More generally, 
if $(\rho^c_i\, \rho_j)^{*} = \xi^\ell$ with  \(\ell=0,1,2,...,N-1\),
\begin{equation}
M_{ij}(q) = a_0\,\hat q^{\ell} + a_1\,\hat q^{\ell + N} +  a_2\,\hat q^{\ell + 2N} + ...\,,  
\label{eq:Mq}
\end{equation}
%
in the vicinity of the symmetric point. 
It crucially follows that the entry $M_{ij}$ is expected to be 
${\cal O}(\epsilon^{\ell})$ whenever ${\rm Im}\,\tau$ is 
sufficiently large 
\footnote{In practice, values of ${\rm Im}\,\tau \cong (2.5 - 3.0)$ 
prove to be sufficiently large (see \cite{Novichkov:2021evw} and further).}.
The power $\ell$ is uniquely determined by how the 
$\Gamma_{\rm N}$ ($\Gamma^\prime_{\rm N}$) representations 
of $\psi^c_i$ and $\psi_j$ decompose under the residual symmetry 
group $Z^T_N$ \cite{Novichkov:2021evw}.
In the case of $A_4$, as we have already indicated, 
$\ell$ can take values $\ell = 0,1,2$.

The discussed result allows, in principle, to obtain fermion mass hierarchies 
without fine-tuning. 
 
%
\section{$A_4$ modular invariant flavor model}
\label{sec:Model}
%
\subsection{Models of quarks}
%

We consider next simple models of quark mass matrices in models with 
 the level $N=3$ modular symmetry ($A_4$ modular flavor symmetry).
 We assign  the $A_4$ representation and the weights for the relevant
 chiral superfields as
 \begin{itemize}
  \item{quark doublet (left-handed) $Q=((u,d),(c,s),(t,b))$\,: $A_4$ triplet with weight $2$,}
  \item{quark singlets (righ-handed) $(d^c,s^c,b^c)$ and
  	$(u^c,c^c,t^c)$\,: $A_4$ singlets $(1',\,1',1')$ with weight $(4,2,0)$, respectively,}
  \item{up and down sector Higgs fields $H_{u,d}=H_{u,d}$\,: $A_4$ singlets $1$ with weight 0,}
\end{itemize}
which are summarized in Table \ref{tab:model}.
\begin{table}[h]
\begin{center}
\renewcommand{\arraystretch}{1.1}
\begin{tabular}{|c|c|c|c|c|} \hline
  & $Q$ & {$(d^c\,,s^c\,,b^c)\,, (u^c\,,c^c\,,t^c)$} &  $H_u$ & $H_d$ \\ \hline
  $SU(2)$ & 2 & 1  & 2 & 2 \\
  $A_4$ & 3 & $(1',\,1',1')$ & $1$ & $1$ \\
  $k$ & 2 & $(4,\,2,\,0)$& 0 & 0 \\ \hline
\end{tabular}
\end{center}
\caption{Assignments of $A_4$ representations and weights in our model.}
\label{tab:model}
\end{table}
Then, the superpotential terms of the  down-type and up-type quark masses are 
written  for the case  in Table \ref{tab:model}:
\begin{align}
 &W_d =
\left[\alpha_d ({\bf Y_3^{(6)}}Q)_{1''} d^c_{1'}
 +\alpha'_d ({\bf Y_{3'}^{(6)}}Q)_{1''} d^c_{1'} 
+\beta_d ({\bf Y_3^{(4)}}Q)_{1''}s^c_{1'} 
+ \gamma_d ({\bf Y_3^{(2)}}Q)_{1''}b^c_{1'}\right] H_d\, ,\nonumber \\
&W_u =
\left[\alpha_u ({\bf Y_3^{(6)}}Q)_{1''} u^c_{1'} 
+\alpha'_u ({\bf Y_{3'}^{(6)}}Q)_{1''} u^c_{1'} 
+\beta_u ({\bf Y_3^{(4)}}Q)_{1''}c^c_{1'} 
+ \gamma_u ({\bf Y_3^{(2)}}Q)_{1''}b^t_{1'}\right] H_u\, ,
\label{superpotential}
\end{align}
where the subscripts of $1',1''$ denote the $A_4$ representations.
The decompositions of the tensor products in $A_4$
are given in Appendix \ref{Tensor}.

  We will also investigate  the possibility 
of right-handed quark assignments 
being $A_4$ singlets $(1',\,1',1')$ 
with weights 
 $(4,2,0)$ for  $(d^c,s^c,b^c)$ and $(6,2,0)$ for  $(u^c,c^c,t^c)$ 
 in Subsection \ref{sub-weight8}.
 In this case modular forms of weight 8 are also involved.

%
\subsection{Modular forms of $A_4$}
%
%
The weight 2 triplet modular forms are given as:
\begin{align}
{\bf Y^{(\rm 2)}_3}
=\begin{pmatrix}Y_1\\Y_2\\Y_3\end{pmatrix}
=\begin{pmatrix}
1+12 q+36 q^2+12 q^3+\dots \\
-6 q^{1/3}(1+7 q+8 q^2+\dots) \\
-18 q^{2/3}(1+2 q+5 q^2+\dots)\end{pmatrix}\,,
\label{Y(2)}
\end{align}
%
where $ q=\exp\,(2\pi i \tau)$.
They satisfy also the constraint \cite{Feruglio:2017spp}:
\begin{align}
Y_2^2+2Y_1 Y_3=0~.
\label{condition}
\end{align}
%
The weight 4, 6 and 8 modular forms of interest 
can be expressed in terms of the weight 2 modular forms.
For the weight 4 modular forms we have:
\begin{align}
&\begin{aligned}
{\bf Y^{(\rm 4)}_1}=Y_1^2+2 Y_2 Y_3 \ , \qquad
{\bf Y^{(\rm 4)}_{1'}}=Y_3^2+2 Y_1 Y_2 \ , \qquad
{\bf Y^{(\rm 4)}_{1''}}=Y_2^2+2 Y_1 Y_3=0 \ , 
\end{aligned}
\nonumber \\
\nonumber \\
&\begin{aligned} {\bf Y^{(\rm 4)}_{3}}=
\begin{pmatrix}
Y_1^{(4)}  \\
Y_2^{(4)} \\
Y_3^{(4)}
\end{pmatrix}
=
\begin{pmatrix}
Y_1^2-Y_2 Y_3  \\
Y_3^2 -Y_1 Y_2 \\
Y_2^2-Y_1 Y_3
\end{pmatrix}\,, 
\end{aligned}
\label{weight4}
\end{align}
%
where ${\bf Y^{(\rm 4)}_{1''}}$ vanishes due to the constraint of 
Eq.\,(\ref{condition}).
%
The expressions for the seven modular forms of weigh 6 read:
\begin{align}
&\begin{aligned}
{\bf Y^{(\rm 6)}_1}=Y_1^3+ Y_2^3+Y_3^3 -3Y_1 Y_2 Y_3  \ , 
\end{aligned} 
\nonumber \\
\nonumber \\
&\begin{aligned} 
{\bf Y^{(\rm 6)}_3}
\equiv 
\begin{pmatrix}
Y_1^{(6)}  \\
Y_2^{(6)} \\
Y_3^{(6)}
\end{pmatrix} 
= (Y_1^2+2 Y_2 Y_3)
\begin{pmatrix}
Y_1  \\
Y_2 \\
Y_3
\end{pmatrix}\ , \qquad
\end{aligned}
\begin{aligned} 
{\bf Y^{(\rm 6)}_{3'}}\equiv
\begin{pmatrix}
Y_1^{'(6)}  \\
Y_2^{'(6)} \\
Y_3^{'(6)}
\end{pmatrix}
=(Y_3^2+2 Y_1 Y_2 )
\begin{pmatrix}
Y_3  \\
Y_1 \\
Y_2
\end{pmatrix}\,. 
\end{aligned}
\label{weight6}
\end{align}
%
We give the expressions also for the nine  modular forms of weight 8:
\begin{align}
&\begin{aligned}
{\bf Y^{(\rm 8)}_1}=(Y_1^2+2Y_2 Y_3)^2 \, , \qquad
{\bf Y^{(\rm 8)}_{1'}}=(Y_1^2+2Y_2 Y_3)(Y_3^2+2Y_1 Y_2)\, , \qquad 
{\bf Y^{(\rm 8)}_{1"}}=(Y_3^2+2Y_1 Y_2)^2\, , 
\end{aligned}  
\nonumber\\
\nonumber\\
&\begin{aligned} 
{\bf Y^{(\rm 8)}_3}
\equiv 
 \hskip -1 mm
\begin{pmatrix}
Y_1^{(8)}  \\
Y_2^{(8)} \\
Y_3^{(8)}
\end{pmatrix} 
 \hskip -1 mm
=(Y_1^2+2 Y_2 Y_3) 
\hskip -1 mm
\begin{pmatrix}
Y_1^2-Y_2 Y_3  \\
Y_3^2 -Y_1 Y_2 \\
Y_2^2-Y_1 Y_3
\end{pmatrix}\,, \ \ \
\end{aligned}
\begin{aligned} 
{\bf Y^{(\rm 8)}_{3'}} \equiv 
 \hskip -1 mm
\begin{pmatrix}
Y_1^{'(8)}  \\
Y_2^{'(8)} \\
Y_3^{'(8)}
\end{pmatrix} 
 \hskip -1 mm
=(Y_3^2+2 Y_1 Y_2 ) 
\hskip -1 mm
\begin{pmatrix}
Y_2^2-Y_1 Y_3 \\
Y_1^2-Y_2 Y_3  \\
Y_3^2 -Y_1 Y_2 
\end{pmatrix}\,.
\end{aligned} 
\label{weight8}
\end{align}
%
 We note that ${\bf Y^{(\rm 8)}_3} = (Y_1^2+2 Y_2 Y_3){\bf Y^{(\rm 4)}_3}$.

At $\tau=i\infty$, the modular forms of interest take very simple forms:
\begin{align}
&\begin{aligned}
{\bf Y^{(\rm 2)}_3}
= Y_0 \begin{pmatrix}
1\\ 
0\\ 
0
\end{pmatrix}\,,
\end{aligned} 
\qquad 
\begin{aligned} 
{\bf Y^{(\rm 4)}_{3}}
= Y_0^2 \begin{pmatrix}
1  \\
0 \\
0
\end{pmatrix}\,, \qquad 
{\bf Y^{(\rm 4)}_1}=Y_0^2\,, \qquad
{\bf Y^{(\rm 4)}_{1'}}=0\,, 
\end{aligned} 
\nonumber\\
\nonumber\\
&
\begin{aligned}
 {\bf Y^{(\rm 6)}_{3}}
=Y_0^3 \begin{pmatrix}
1\\
0\\ 
0
\end{pmatrix}\,,
\end{aligned} 
\qquad
\begin{aligned} 
{\bf Y^{(\rm 6)}_{3'}}=0\,, \qquad 
{\bf Y^{(\rm 6)}_1}= Y_0^3\,,
\end{aligned}
\nonumber\\
\nonumber \\
&
\begin{aligned} 
{\bf Y^{(\rm 8)}_{3}} 
= Y_0^4 \begin{pmatrix}
1  \\
0 \\
0
\end{pmatrix}\,,
\end{aligned} \qquad 
\begin{aligned}
{\bf Y^{(\rm 8)}_{3'}}=0\,, \quad
{\bf Y^{(\rm 8)}_1}=Y^4_0\,, \quad {\bf Y^{(\rm 8)}_{1'}}=0\,, 
\quad 
{\bf Y^{(\rm 8)}_{1''}}= 0\,,
\label{tauiinf}
\end{aligned} 
\end{align}
%
where we can take a normalization $Y_0=1$.

%
\subsection{Modular forms  ``close'' to $\tau=i\infty$}
\label{Modular}
%

We present next  the behavior of modular forms 
``close'' to $\tau=i\infty$. By ``close to'', 
or ``in the vicinity'' of, $\tau=i\infty$ 
we mean values of ${\rm Im}\,[\tau]$ 
which are sufficiently large, 
e.g., ${\rm Im}\,[\tau] \sim 2.5-3.0$.
As we will see, values of  ${\rm Im}\,[\tau] \sim 2.5$ 
are also typically required to fit the quark mass and mixing data.
We find first the expressions of the modular forms $Y_1(\tau)$, $Y_2(\tau)$ 
and $Y_3(\tau)$ in Eq.\,\eqref{Y(2)} close to  $\tau=i\infty$.
We use the small real  parameter $\epsilon$ as given in Eq.\,\eqref{eq:q}, and 
 the phase $p$ 
in terms of $\tau$ as:
\begin{align}
\begin{aligned}
\epsilon=\exp\left (-\frac23\pi\, {\rm Im}\,[\tau]\right ) \,, \qquad
p=\exp\left (\frac23\pi\,i\, {\rm Re}\, [\tau]\right )\,, 
\end{aligned}
\label{ep}
\end{align}
%
where  $\epsilon\ll 1$.
 At  ${\rm Im}\,[\tau] = 2.5$, for example, 
$\epsilon \cong 5.322\times 10^{-3}$, $\epsilon^2 \cong 2.832 \times 10^{-5}$, 
$\epsilon^3 \cong 1.507\times 10^{-7}$, 
$\epsilon^4 \cong 8.020\times 10^{-10}$, 
$\epsilon^5 \cong 4.268\times 10^{-12}$
and $\epsilon^6 \cong 2.271\times 10^{-14}$.
%
The expansion parameter $q$ in Eq.\,\eqref{Y(2)} is written 
in terms of $\epsilon$ and $p$ as: 
\begin{align}
q\equiv\exp \,(2i\pi\tau) =(p\,\epsilon)^3\,.
\label{q}
\end{align}
%
For the components of 
triplet weight 2 modular form ${\bf Y^{(\rm 2)}_3}$
we obtain in terms of the small parameter $(p\,\epsilon)$:
\begin{align}
&Y_1(\tau)\simeq 1+ 12\,(p\,\epsilon)^3 + 36\,(p\,\epsilon)^6 + 
12\,(p\,\epsilon)^9 + {\cal O}(\epsilon^{12})\,,
\nonumber
\\
&Y_2(\tau)\simeq -6 \, (p\,\epsilon) - 42\,(p\,\epsilon)^4 
- 48\,(p\,\epsilon)^7 + {\cal O}(\epsilon^{10})\,, 
\nonumber
\\ 
&Y_3(\tau)\simeq -18 \,(p\,\epsilon)^2 -36\,(p\,\epsilon)^5 
-\,90\,(p\,\epsilon)^8 +  {\cal O}(\epsilon^{11})\,.
\label{ep2}
\end{align}
%


It follows from Eq.\,(\ref{superpotential}) that only triplet 
modular forms will enter into the expressions of the 
down-type and up-type quark mass matrices. 
We use the results in Eq.\,(\ref{ep2}) 
and Eqs.\,(\ref{weight4}), (\ref{weight6}) and (\ref{weight8})
to express the  higher weight triplet modular forms $ Y_i^{(k)}$
in terms of $(p\,\epsilon)$. 
For the   weight $4$ triplet modular form we get:
{\bf 
\begin{align}  
&Y_1^{(4)}(\tau)\simeq 1 - 84\,(p\,\epsilon)^3  
- 756\,(p\,\epsilon)^6  + {\cal O}(\epsilon^9)\, 
\simeq  1 - 84\,(p\,\epsilon)^3  +{\cal O}(10^{-11})\,,
\nonumber 
\\
&Y_2^{(4)}\simeq 6\,(p\,\epsilon) + 438\,(p\,\epsilon)^4 + 
2016\,(p\,\epsilon)^7 + {\cal O}(\epsilon^9)\,
\simeq  6\,(p\,\epsilon) + {\cal O}(10^{{-7}})\,, 
\nonumber 
\\ 
&Y_3^{(4)}\simeq 54\,(p\,\epsilon)^2 + 756\,(p\,\epsilon)^5 + 
3530\,(p\,\epsilon)^8 +  {\cal O}(\epsilon^{11}) 
\simeq  54\,(p\,\epsilon)^2 + {\cal O}(10^{-9})\,.
\label{ep4}
\end{align}
}
%
In the case of the two weight $6$ triplet modular forms 
the expressions read:
{\bf 
\begin{align} 
&Y_1^{(6)}(\tau) \simeq 1 + 252\,(p\,\epsilon)^3  + 5076\,(p\,\epsilon)^6 
+ 41292\,(p\,\epsilon)^9 + {\cal O}(\epsilon^{12}) \nonumber\\ 
&\qquad \quad\simeq 1 + 252\,(p\,\epsilon)^3  + {\cal O}(10^{-10})\,,
\nonumber
\\
&Y_2^{(6)}\simeq -\,6\,(p\,\epsilon) -\, 1482\,(p\,\epsilon)^4 
- 11088\,(p\,\epsilon)^7 + {\cal O}(\epsilon^{10})
\simeq   -\,6\,(p\,\epsilon) -\, {\cal O}( 10^{-5})\,,
\nonumber
\\ 
&Y_3^{(6)} \simeq -\, 18\,(p\,\epsilon)^2 -\, 4356\,(p\,\epsilon)^5 
-\, 47610\,(p\,\epsilon)^8 + {\cal O}(\epsilon^{11})
\simeq  -\, 18\,(p\,\epsilon)^2 -\,{\cal O}(10^{-8})\,,
\nonumber
\\
&Y_1^{'(6)}(\tau) \simeq 216\,(p\,\epsilon)^3 - 1296\,(p\,\epsilon)^6 
+ 1944\,(p\,\epsilon)^9 + {\cal O}(\epsilon^{12}) 
\simeq 216\,(p\,\epsilon)^3 -\, {\cal O}(10^{-11})\,,
\nonumber
\\ 
&Y_2^{'(6)}(\tau) \simeq -12\,(p\,\epsilon) 
-\, 40\,(p\,\epsilon)^4 + 480\,(p\,\epsilon)^7 +  {\cal O}(\epsilon^{10})\,
\simeq -12\,(p\,\epsilon) -\, {\cal O}(10^{-8})\,,
\nonumber
\\ 
&Y_3^{'(6)}(\tau)\simeq 72\,(p\,\epsilon)^2 -  72\,(p\,\epsilon)^5 
- 2016\,(p\,\epsilon)^8 + {\cal O}(\epsilon^{11})\,
\simeq 72\,(p\,\epsilon)^2 + {\cal O}(10^{-11})\,.
\label{ep6}
\end{align}
}
%
Finally we give the expansions for the two triplet weight $8$  modular forms:
{\bf 
\begin{align} 
&Y_1^{(8)}(\tau) \simeq 1 + 156\,(p\,\epsilon)^3  - 18758\,(p\,\epsilon)^6 
-\,383864\,(p\,\epsilon)^9 + {\cal O}(\epsilon^{12})\nonumber\\
&\qquad \quad\simeq 1 + 156\,(p\,\epsilon)^3 +\,  {\cal O}(10^{-10})\,, 
\nonumber
\\
&Y_2^{(8)}\simeq 6\,(p\,\epsilon) + 1878\,(p\,\epsilon)^4 
+  120144\,(p\,\epsilon)^7 + {\cal O}(\epsilon^{10}) \nonumber\\
&\qquad\simeq 6\,(p\,\epsilon) +\,{\cal O}(10^{-6})\,,
\nonumber
\\ 
&Y_3^{(8)} \simeq 54\,(p\,\epsilon)^2 + 13716\,(p\,\epsilon)^5 
+ 301610\,(p\,\epsilon)^8 +  {\cal O}(\epsilon^{11}) \nonumber\\
&\qquad \simeq 54\,(p\,\epsilon)^2 +\, {\cal O}(10^{-7})\,,
\nonumber
\\
&Y_1^{'(8)}(\tau) \simeq -\,648\,(p\,\epsilon)^3 - 3888\,(p\,\epsilon)^6 
+ 17256\,(p\,\epsilon)^9 +  {\cal O}(\epsilon^{12}) \nonumber\\
&\qquad \quad\ \simeq -\,648\,(p\,\epsilon)^3 -\,{\cal O}(10^{-10})\,, 
\nonumber
\\ 
&Y_2^{'(8)}(\tau) \simeq -\,12\,(p\,\epsilon) + 1104\,(p\,\epsilon)^4
+ 768\,(p\,\epsilon)^7  + {\cal O}(\epsilon^{10})\nonumber\\
&\qquad\quad  \simeq  -\,12\,(p\,\epsilon) +\,{\cal O}(10^{-6})\,,
\nonumber
\\ 
&Y_3^{'(8)}(\tau) \simeq -\,72\,(p\,\epsilon)^2 -  4680\,(p\,\epsilon)^5 
+ 15840\,(p\,\epsilon)^8 + {\cal O}(\epsilon^{11}) \nonumber\\
&\qquad \quad\simeq -\,72\,(p\,\epsilon)^2 -\, {\cal O}(10^{-8})\,.
\label{ep8}
\end{align}
}
%
In Eqs.\,(\ref{ep4}) - (\ref{ep8})
the numerical estimates of the order of the corrections  
are for  ${\rm Im}\,[\tau] = 2.5$ and 
we have kept the terms in the expansions which 
are not smaller than $\sim 10^{-5}$.

%
\subsection{The Quark mass matrices}
%
%

We obtain  the quark mass matrices from the superpotential
in Eq.\,\eqref{superpotential}.
In the right-left (RL) convention they have the form:
\begin{align}
&  M_d =v_d
\begin{pmatrix}
\alpha'_d  & 0 & 0  \\
0 & \beta_d & 0 \\
0& 0 & \gamma_d \\
\end{pmatrix}
\begin{pmatrix}
\tilde Y_3^{(6)} &  \tilde Y_2^{(6)} &  \tilde Y_1^{(6)} \\
Y_3^{(4)} & Y_2^{(4)} &Y_1^{(4)} \\
Y^{(2)}_3 & Y^{(2)}_2 &Y^{(2)}_1 
\end{pmatrix},\quad
M_u =v_u
\begin{pmatrix}
\alpha'_u  & 0  & 0  \\
0 & \beta_u & 0 \\
0& 0 & \gamma_u \\
\end{pmatrix}
\begin{pmatrix}
\tilde Y_3^{(6)} &  \tilde Y_2^{(6)} &  \tilde Y_1^{(6)} \\
Y_3^{(4)} & Y_2^{(4)} &Y_1^{(4)} \\
Y^{(2)}_3 & Y^{(2)}_2 &Y^{(2)}_1 \\
\end{pmatrix},
\label{Massmatrix-I}
\end{align}
where $Y_i^{(2)}\equiv Y_i$ and $v_{d\,(u)}$ denotes the vacuum expectation value of $H_{d\,(u)}$, and 
\begin{align}
\tilde Y_i^{(6)}=  g_{q} Y_i^{(6)}+  \, Y_i^{'(6)}, \qquad 
g_{q}\equiv\alpha_q/\alpha'_q\,\qquad (i=1,\,2,\,3,\quad q=d,\,u)\, .
\end{align}

We take the modular invariant  kinetic terms simply by
{\footnote{Possible non-minimal additions to the K\"ahler potential, 
		compatible with the modular symmetry, may jeopardise the predictive power 
		of the approach~\cite{Chen:2019ewa}. 
		This problem is the subject of ongoing research.} }
\begin{equation}
\sum_I\frac{|\partial_\mu\psi^{(I)}|^2}{\langle-i\tau+i\bar{\tau}\rangle^{k_I}} ~,
\label{kinetic}
\end{equation}
%
where $\psi^{(I)}$ denotes a chiral superfield with weight $k_I$,
and $\bar\tau$ is the anti-holomorphic modulus.
{Since  the (anti-)holomorphic modulus is a dynamical field,
   it becomes a complex number after the modulus $\tau$ takes a VEV.
Then, one can set $\bar \tau=\tau^*$.}

It is important to  address  the transformation needed to get the kinetic 
terms of matter superfields in canonical form because the
terms  in  Eq.\,(\ref{kinetic}) are not canonical. 
Therefore, we normalize the superfields as:
\begin{eqnarray}
\psi^{(I)}\rightarrow  \sqrt{(2{\rm Im}\tau_q)^{k_I}} \, \psi^{(I)}\,.
\label{canonical}
\end{eqnarray}
%
The canonical form  is obtained by an overall normalization, which
shifts our parameters  such as:
\begin{eqnarray}
&&\alpha_q \rightarrow \hat\alpha_q= \alpha_q\, \sqrt{(2 {\rm Im} \tau_q)^{6} }
  =\alpha_q\,(2 {\rm Im} \tau_q)^{3},\, \nonumber\\
&&\alpha'_q \rightarrow \hat\alpha'_q= \alpha'_q\, \sqrt{(2 {\rm Im} \tau_q)^{6} }
=\alpha'_q\,(2 {\rm Im} \tau_q)^{3}\, ,\nonumber\\
&&\beta_q  \rightarrow \hat\beta_q = \beta_q  \, \sqrt{(2 {\rm Im} \tau_q)^{4} }
= \beta_q  \, (2 {\rm Im} \tau_q)^2\, , \nonumber\\
&&\gamma_q  \rightarrow \hat\gamma_q = \gamma_q \sqrt{(2 {\rm Im} \tau_q)^{2} }
=\gamma_q \, (2 {\rm Im} \tau_q)\, . 
\label{shift}
\end{eqnarray}

%
\subsection{Quark mass matrices  at  $\tau=i\infty$}
%
%

Since  $\epsilon=0$ at $\tau=i\infty$,
the modular forms $Y_2^{(k)}$ and $Y_3^{(k)}\,(k=2,4,6)$ vanish.
Then, the quark mass matrices are given by:
\begin{align}
\begin{aligned}
{ M^{(0)}_q} =v_q
\begin{pmatrix}
g_{q}\, \hat\alpha'_q  & 0 & 0 \\
0 & \hat\beta_q & 0\\
0 & 0 &\hat\gamma_q
\end{pmatrix} 
\begin{pmatrix}
0& 0 & 1 \\
0 &0 &  1\\
0 & 0& 1
\end{pmatrix}_{RL} \, , \qquad q=d\,,u\,.
\end{aligned}
\label{quark}
\end{align}
%
 These are  rank one matrices. Thus
  we have two massless down-type quarks
and two massless up-type quarks
  at the fixed point $\tau=i\infty$.
The  matrices ${ M^{(0)\dagger}_q  M^{(0)}_q}$, $q=d,\,u$,
whose eigenvalues are the squared quark masses,
are given by:

\begin{align}
{\cal M}_q^{2(0)}\equiv  { M^{(0)\dagger}_q M^{(0)}_q} 
= v_q^2
\begin{aligned}
\begin{pmatrix}
0&0 & 0 \\
0&0 &   0  \\
0 & 0 & |g_{q}|^2\,|\hat\alpha'_q|^2 + |\hat\beta_q|^2 + |\hat\gamma_q|^2 \\
\end{pmatrix}
\end{aligned}\,.
\label{}
\end{align}
%
%
\subsection{Quark mass matrices close to $\tau=i\infty$ and mass hierarchy}    
\label{Quark642}
%

The quark mass matrices in Eq.\,\eqref{quark} 
are corrected due to the deviation from  $\tau=i\infty$.
By using  modular forms  of weight 2, 4 and 6
in Eqs.\,\eqref{ep2}, \eqref{ep4}, \eqref{ep6}, we obtain
the deviation from  
${ M_q^{(0)}}$ given in Eq.\,(\ref{quark}).
The correction is expressed in terms of the small variable $\epsilon$ and
the phase $p$ defined in Eq.\,\eqref{ep}
In the 
leading order approximation in $\epsilon$,
${ M_q}$ is given by:
 \begin{align}
{ M}_q
=v_q
\begin{pmatrix}
 \hat\alpha'_q  & 0 & 0 \\
0 & \hat\beta_q & 0\\
0 & 0 &\hat\gamma_q
\end{pmatrix} 
\begin{pmatrix}
18\,(\epsilon\,p)^2 (4-g_q)& -6\,(\epsilon\,p) (2+g_{q}) &  g_{q} \\
54\,(\epsilon\,p)^2& 6\,(\epsilon\,p) &   1  \\
	-18\,(\epsilon\,p)^2 &-6\,(\epsilon\,p)&1
\end{pmatrix}\,.
\label{Massmatrix}
\end{align}
%
 Correspondingly, 
${\cal M}_q^{2}\equiv  { M_q^\dagger M_q}$ has the following structure: 
\begin{align}
{\cal M}_q^2  \sim
\begin{pmatrix} 
\epsilon^4 & \epsilon^{3} p^{*} & \epsilon^{2} p^{*2} \cr  
\epsilon^{3} p  &\epsilon^2&\epsilon p^*\cr
\epsilon^2 p^2  &\epsilon p& 1
\end{pmatrix}\, .
\label{matrix-app}
\end{align}
%

Using Eq.\,(\ref{Massmatrix}), we obtain the 
elements of
${\cal M}_q^{2}$ in leading order in $\epsilon$:
\begin{align}
&{\cal M}_q^{2} [1,1]=
 324\,\epsilon^4 \left [|\hat\gamma_q|^2+
9|\hat\beta_q|^2 + |\hat\alpha'_q|^2|g_q - 4|^2\right ]\,,\nonumber\\
&{\cal M}_q^{2} [2,2]=
36\,\epsilon^2 \left [|\hat\gamma_q|^2+
|\hat\beta_q|^2 + |\hat\alpha'_q|^2|g_q + 2|^2\right ]\,,\nonumber\\
&{\cal M}_q^{2} [3,3]=
|\hat\gamma_q|^2+
|\hat\beta_q|^2 + |\hat\alpha'_q|^2|g_q|^2\,,\nonumber\\
&{\cal M}_q^{2} [1,2]= 108\,\epsilon^3\,p^* \left [|\hat\gamma_q|^2+
3|\hat\beta_q|^2 + |\hat\alpha'_q|^2(g_q+2)(g^*_q-4)\right ]\,,\nonumber\\
&{\cal M}_q^{2} [1,3]= -18\,\epsilon^2 \, p^{*2} \left [|\hat\gamma_q|^2-
3|\hat\beta_q|^2 + |\hat\alpha'_q|^2 g_q(g^*_q-4)\right ]\,,\nonumber\\
&{\cal M}_q^{2} [2,3]=-6\,\epsilon\, p^{*} \left [|\hat\gamma_q|^2-
|\hat\beta_q|^2 + |\hat\alpha'_q|^2 g_q(g^*_q+2)\right ]\,,\nonumber\\
&{\cal M}_q^{2} [2,1]={\cal M}_q^{2} [1,2]^*\,,\qquad
{\cal M}_q^{2} [3,1]={\cal M}_q^{2} [1,3]^*\,,\qquad
{\cal M}_q^{2} [3,2]={\cal M}_q^{2} [2,3]^*\,,
\label{masscomponent}
\end{align}
%
where the elements are given in units of $v_q^2$.
For the determinant of ${\cal M}_q^2$ we find:
\begin{align}
{\rm Det}\,[{\cal M}_q^2]=m^2_{q_1}m^2_{q_2}m^2_{q_3} =(12)^6|\hat\alpha'_q|^2|\hat\beta_q|^2 
|\hat\gamma_q|^2\, v_q^6\, \epsilon^6\,,
\label{eq:Det}
\end{align}
{with $d_1\equiv d$, $d_2\equiv s$, $d_3\equiv b$, $u_1\equiv u$, $u_2\equiv c$ and $u_3\equiv t$.}
Thus, ${\rm Det}\,[{\cal M}_q^2]$ is $g_q$ independent. 
 The heaviest quark mass  $m_{q_3}$ can be obtained
  to a good approximation from $m_{q_3}^2 \cong {\rm Tr}({\cal M}_q^2)$.
  We find:
\begin{align}
m_{q_3}^2\simeq  (|\hat\alpha'_q|^2|g_q|^2 + |\hat\beta_q|^2 
+ |\hat\gamma_q|^2) v_q^2 \,.
\label{eq:mq3e2}
\end{align}
%
 It is also not difficult to obtain an expression for $m_{q_2}^2 m_{q_3}^2$, 
which to a good approximation is given by the determninant of the
2-3 sector:
\begin{align}
m_{q_2}^2 m_{q_3}^2 \simeq (12)^2 \epsilon^2
(|\hat\alpha'_q|^2 |\hat\beta_q|^2 |g_q+1|^2 + 
|\hat\alpha'_q|^2 |\hat\gamma_q|^2 
+ |\hat\beta_q|^2|\hat\gamma_q|^2)v_q^4 \,.
\label{subdet}
\end{align}
%

Suppose  that $\alpha'_q$, $\beta_q$ and $\gamma_q$,
which appear in the superpotential in Eq.\,\eqref{superpotential} 
are in magnitude of the same order.
As follows from  Eq.\,\eqref{shift}, 
 $\hat\alpha'_q$, $\hat\beta_q$ and $\hat\gamma_q$ are enhanced by 
the factors associated with the renormalisation of the matter fields 
(see Eqs.\,(\ref{kinetic}) and  (\ref{canonical})).   
These factors are powers of  $2 {\rm Im}\,\tau$, which 
for $\tau\rightarrow i\infty$ has a arelatively large value 
(in the case of interest,  ${\rm Im}\,\tau \sim 2.5$).
After taking into account that
for $|\alpha'_q| \sim |\beta_q| \sim |\gamma_q|$, as it follows from 
Eq.\,\eqref{shift}, we have 
$|\hat\alpha'_q|^2 \gg |\hat\beta_q|^2 \gg |\hat\gamma_q|^2$,
we get from Eqs.\,(\ref{eq:Det}) - (\ref{subdet}):
\begin{align}
&m_{q_3} \simeq \hat\alpha'_q\, g_q=\alpha'_q I_\tau^3\, g_q\,,~\nonumber\\
&m_{q_2} \simeq \frac{g_q + 1}{g_q} \hat\beta_q (12\,\epsilon)=
\frac{g_q + 1}{g_q}\beta_q I_\tau^2 (12\,\epsilon)\,,~\nonumber\\
&m_{q_1} \simeq \frac{1}{g_q}\hat\gamma_q (12\,\epsilon)^2
=\frac{1}{g_q}\gamma_q I_\tau (12\,\epsilon)^2\,,
\end{align}
%
where $I_\tau \equiv 2{\rm Im}\,\tau$ and we have 
assumed for simplicity that the constant $g_q$ is real and 
that $g_q \gtap 1$.
Thus, for the mass ratios we obtain: 
\begin{align}
m_{q_3} : m_{q_2}: m_{q_1}\simeq
I_\tau^3 g_q: (12\,\epsilon) I_\tau^2 \frac{g_q+1}{g_q}: 
(12\,\epsilon)^2I_\tau \frac{1}{g_q}
= I_\tau^3 g_q \left[
1: \left(\frac{12\epsilon}{I_\tau g_q}\frac{g_q+1}{g_q}\right): 
\left(\frac{12\epsilon}{I_\tau g_q}\right)^2 \right ]\,.
\label{mass-ratios}
\end{align}
%
This result implies that in the case of 
 $|\alpha'_q| \sim |\beta_q| \sim |\gamma_q|$
and $|g_q|\sim 1$ in the considered model,  
the quark mass hierarchies are determined by 
${\rm Im}\,\tau$ (we recall that 
$\epsilon = {\rm exp}(-\,2\pi {\rm Im}\,\tau/3)$).
The constant  $|g_q|$ can play a role in 
obtaining the correct quark mass ratios if, e.g., 
$|g_q| >> 1$.

 Indeed,
as we show below, both down-type  and     
up-type quark mass hierarchies can be explained with a common  
$ \tau$ in the down-type and up-type quark matrices  
with  ${\rm Im}\tau \sim 2.5$, 
by taking $|g_d|\sim 1$ and $|g_u|\sim 10$.
We will explore also the  alternative possibility of having two different 
moduli in the down-type and up-type quark mass matrices $\tau_d$  
and $\tau_u$ and thus two different small parameters 
$\epsilon_d$ and $\epsilon_u$, $\epsilon_d \neq \epsilon_u$. 

We note finally that  
the unitary matrices $ U_q\, (q=d,\,u)$, which diagonalise the matrix
${ M_q^\dagger M_q}$ as
${ M_q^\dagger M_q = U_q\, {\rm diag}(m^2_{q_1},m^2_{q_2},m^2_{q_3})\,U^\dagger_q}$,
with $u_1\equiv u$, $u_2\equiv c$, $u_3\equiv t$,
$d_1\equiv d$, $d_2\equiv s$ and $d_3\equiv b$,
form the CKM quark mixing matrix: ${ U_{\rm CKM} = U^\dagger_u\, U_d}$.
We use the parametrisation of ${ U_{\rm CKM}}$, 
and thus the convention for three quark mixing angles 
and CPV phase present in ${ U_{\rm CKM}}$,
$\theta_{12}$, $\theta_{23}$, $\theta_{13}$ and $\delta$,
proposed in \cite{ParticleDataGroup:2018ovx}.
%

%
\section{Input data of quark masses, CKM elements}
\label{sec:inputs} 
%

 The  modulus $\tau$ breaks the modular invariance
by obtaining a VEV at some high mass scale.
We assume this to be the GUT scale. Correspondingly, the values of the
quark masses and CKM parameters at the GUT scale play the role of
the observables that have to be reproduced by the considered quark flavour
model. They are obtained using the renormalisation group
(RG) equations which describe the ``running'' of the observables of interest
from the electroweak scale, where they are measured, to the GUT scale.
In the analyses which follow we adopt the numerical values
of the quark Yukawa couplings 
at the GUT scale $2\times 10^{16}$ GeV
derived in the framework of the minimal SUSY breaking scenarios
with $\tan\beta=5$ 
\cite{Antusch:2013jca}:
\begin{align}
\begin{aligned}
&\frac{y_d}{y_b}=9.21\times 10^{-4}\ (1\pm 0.111) \, , 
\qquad
\frac{y_s}{y_b}=1.82\times 10^{-2}\ (1\pm 0.055) \, , \\
\rule[15pt]{0pt}{1pt}
&\frac{y_u}{y_t}=5.39\times 10^{-6}\ (1\pm 0.311)\,  , 
\qquad
\frac{y_c}{y_t}=2.80\times 10^{-3}\ (1\pm 0.043)\,. \\
\end{aligned}\label{Datamass}
\end{align}
%
The quark masses are given as $m_q=y_q v_H$ with $v_H=174$ GeV.
The choice of relatively small value of $\tan\beta$ allows us to 
avoid relatively large $\tan\beta$-enhanced threshold corrections in 
the RG running of the Yukawa couplings. 
We set these corrections to zero.

  Assuming that both the ratios of the down-type and up-type 
quark masses, $m_{b}, m_{s}, m_{d}$ and $m_{t}, m_{c}, m_{u}$, 
are determined in the model by the small parameter 
$\hat\epsilon$, 
{which is defined in the comparison of the following equations 
with in Eq.\,\eqref{mass-ratios}}, \\
{\bf which is defined in the following equations, to be compared with 
Eq.\,\eqref{mass-ratios},}\\
we have:
\begin{align}
\label{down-hierarchy}
&m_{b} : m_{s}: m_{d}\simeq 1:\hat\epsilon:\hat\epsilon^2\,,~~
\hat\epsilon = 0.02\sim 0.03\,,\\[0.25cm] 
&m_{t} : m_{c}: m_{u} \simeq 1:\hat\epsilon:\hat\epsilon^2\,,~~
\hat\epsilon =0.002\sim 0.003\,.
\label{up-hierarchy}
 \end{align}
%
Thus, the required 
$\hat\epsilon$ for the 
description of the down-type and up-type quark 
mass hierarchies differ approximately by one order of 
magnitude. As indicated by Eq.\,\eqref{mass-ratios}, 
this inconsistency can be ``rescued'' by relaxing the 
requirement on the constant $|g_u|$ in the up-quark sector,  
such as $|g_u|={\cal O}(1)\rightarrow {\cal O}(10)$ leading to
\begin{align}
%
m_{t} : m_{c}: m_{u} \simeq 
1 : \frac{\hat\epsilon}{|{g_u|}}: 
\left (\frac{\hat\epsilon}{{|g_u|}}\right )^2\,,
\end{align}
%
with $\hat\epsilon = 0.02\sim 0.03$.

{The quark flavor mixing is given by the CKM matrix,
	which has  three independent mixing angles and one CP violating phase.
  These mixing angles are given by 
  the absolute values of the following three CKM elements.
We take the present data on the three CKM elements
 in Particle Data Group (PDG) edition 
of Review of Particle Physics 
\cite{ParticleDataGroup:2022pth} as:}
\begin{align}
 \begin{aligned}
 |V_{us}^{\rm }|=0.22500\pm 0.00067 \, , \quad
 |V_{cb}^{\rm }|=0.04182^{\pm 0.00085}_{-0.00074} \,,  \quad
 |V_{ub}^{\rm }|=0.00369\pm 0.00011\, .
 \end{aligned}\label{DataCKM}
 \end{align}
%
By using these values as input and $\tan\beta=5$ we obtain 
the CKM 
mixing angles  at the GUT scale of $2\times 10^{16}$ GeV
\cite{Antusch:2013jca}:
  \begin{align}
 |V_{us}^{\rm }|=0.2250\,(1\pm 0.0032) \, , \quad
|V_{cb}^{\rm }|=0.0400\, (1\pm 0.020) \,,  \quad
 |V_{ub}^{\rm }|=0.00353\,(1\pm 0.036)\, .
\label{DataCKM-GUT}
 \end{align}
%
%
The tree-level decays of $B\to D^{(*)}K^{(*)}$ are used as the standard candle
of the CP violation.  The CP violating phase of latest world average  
is given in PDG2022 \cite{ParticleDataGroup:2022pth} as:
\begin{equation}
\delta_{CP}={66.2^\circ}^{+ 3.4^\circ}_{-3.6^\circ}\,. 
\label{CKMphase}
\end{equation}
%
Since the phase is almost independent of the evolution of RGE's,
we refer to this value in the numerical discussions.
The rephasing invariant CP violating measure $J_{\rm CP}$ \cite{Jarlskog:1985ht}
is also given in \cite{ParticleDataGroup:2022pth}:
\begin{equation}
J_{\rm CP}=3.08^{+0.15}_{-0.13} \times 10^{-5} \,.
\label{JCP}
\end{equation}
%
Taking into account the RG effects on the mixing angles 
for $\tan\beta = 5$, we have at the GUT scale $2\times 10^{16}$ GeV: 
\begin{equation}
J_{\rm CP}= 2.80^{+0.14}_{-0.12}\times 10^{-5}\,.
\label{JCPGUT}
\end{equation}
%

%
\section{Numerical Analyses}
\label{sec:Numerical}
%
%
 In present Section we show results 
of the numerical analyses of the considered $A_4$ quark flavour model.
We perform a fit of the quark masses, CKM mixing angles 
and CP violating phase $\delta_{\rm CP}$ at the GUT scale, 
whose values are given in Section \ref{sec:inputs}. 

{ More specifically, we  perform a systematic scan restricting 
model parameters in the relevant ranges 
to reproduce the values of the observables 
by using a measure of goodness of the fit defined in 
Appendix \ref{fit}, as shown in the following subsections.}
Several phenomenological possibilities are investigated, 
as specified below.

In the considered model we have four constant parameters in 
each of the two quark sectors 
$\alpha_q$, $\alpha'_q$, $\beta_q$ and $\gamma_q$, $q=d, \, u$.
We are interested first of all whether it is possible to describe 
the down-type and up-type quark mass hierarchies 
in terms of powers of the small parameter
$\epsilon = {\rm exp}(-2\pi {\rm Im}\,\tau/3)$ 
avoiding fine-tuning of the constants present in the model.
Thus, 
except possibly for $|g_u| = |\alpha_u/\alpha'_u|$,    
all other constants should be of the same order in magnitude, i.e.,
$|\beta_q|/|\alpha'_q| \sim |\gamma_q|/|\alpha'_q| \sim {\cal O}(1)$, $q=d,\, u
$, 
and $|g_d| = |\alpha_d|/|\alpha'_d| \sim {\cal O}(1)$.     
In this case their influence on the strong quark 
mass hierarchies of interest is insignificant 
\cite{Novichkov:2021evw}. 
The minimal number of parameters in the model corresponds 
the case of real constants 
$\alpha_q$, $\alpha'_q$, $\beta_q$ and $\gamma_q$, $q=d,\, u$.
The reality of the constants can be ensured by 
imposing the condition of exact  gCP 
symmetry in the model \cite{Novichkov:2019sqv}. The gCP 
symmetry will be broken by the complex value of $\tau$  
\footnote{In order for the gCP symmetry to be broken 
the value of $\tau$ should not lie on the 
border of the fundamental domain of the modular group 
and ${\rm Re}(\tau) \neq 0$ \cite{Novichkov:2021evw}.
}.
It can be broken also by some, or all, constants being complex 
and we will analyse also these cases.

%
\subsection{Quark mass hierarchies with common $\tau$
  in ${\cal M}_d$ and ${\cal M}_u$}

 In the present subsection we investigate the case that the
  down-type and up-type quark mass matrices depend on the same modulus
  $\tau$. In what concerns the violation of the CP symmetry, there are
  four phenomenological possibilities, which we consider below.

%
\subsubsection{ Real $g_d$ and $g_u$}
\label{CKMreal}
%

 In this case the CP symmetry is violated only by the VEV of the modulus $\tau$.
A tentative sample set for the  fitting of the 
values of the observables at the GUT scale given in Section \ref{sec:inputs}
and the results obtained are  
presented in Tables \ref{tab:input2} and  \ref{tab:output2}.
For the expansion parameter we get 
$\epsilon \simeq 5.8\times 10^{-3}$.

\begin{table}[h]
	\begin{center}
		\renewcommand{\arraystretch}{1.1}
		\begin{tabular}{|c|c|c|c|c|c|c|} \hline
			\rule[14pt]{0pt}{3pt}  
			$\tau$ & $\frac{\beta_d}{\alpha'_d}$ & $\frac{\gamma_d}{\alpha'_d}$ & $g_d$ 
			& $\frac{\beta_u}{\alpha'_u}$ & $\frac{\gamma_u}{\alpha'_u}$ & $g_u$
			\\
			\hline
			\rule[14pt]{0pt}{3pt}  
			$-0.493+i\,2.459$   &$3.33$ & $0.98$  & $-0.70$  & $2.43$
			& $1.07$&  $-11.8$\\ \hline
		\end{tabular}
	\end{center}
	\caption{ Values of the constant parameters obtained in the fit of the 
		quark mass ratios and of 
		CKM mixing angles. 
See text for details.
	}
	\label{tab:input2}
\end{table}

\begin{table}[h]
	\begin{center}
		\renewcommand{\arraystretch}{1.1}
		\begin{tabular}{|c|c|c|c|c|c|c|c|c|} \hline
			\rule[14pt]{0pt}{3pt}  
			& $\frac{m_s}{m_b}\hskip -1 mm\times\hskip -1 mm 10^2$ 
			& $\frac{m_d}{m_b}\hskip -1 mm\times\hskip -1 mm 10^4$& $\frac{m_c}{m_t}\hskip -1 mm\times\hskip -1 mm 10^3$&$\frac{m_u}{m_t}\hskip -1 mm\times\hskip -1 mm 10^6$&
			$|V_{us}|$ &$|V_{cb}|$ &$|V_{ub}|$&$J_{\rm CP}$
			\\
			\hline
			\rule[14pt]{0pt}{3pt}  
			Fit &$1.56$ & $ 6.43$
			& $2.65$ & $1.67 $&
			$0.2246$ & $0.0787$ & $0.00366$ &
			$2.0\hskip -1 mm\times\hskip -1 mm 10^{-10}$
			\\ \hline
			\rule[14pt]{0pt}{3pt}
			Exp	 &$1.82$ & $9.21$ 
			& $2.80$& $ 5.39$ &
			$0.2250$ & $0.0400$ & $0.00353$ &$2.8\hskip -1 mm\times\hskip -1 mm 10^{-5}$\\
			$1\,\sigma$	&$\pm 0.10$ &$\pm 1.02$ & $\pm 0.12$& $\pm 1.68$ &$ \pm 0.0007$ &
			$ \pm 0.0008$ & $ \pm 0.00013$ &$^{+0.14}_{-0.12}\hskip -1 mm\times \hskip -1 mm 10^{-5}$\\ \hline  
		\end{tabular}
	\end{center}
	\caption{Results of the fit of the quark mass ratios and 
		CKM mixing angles.  $J_{\rm CP}$ factor is output.
		'Exp' denotes the respective values at the GUT scale, 
		including $1\sigma$ errors
		.
	}
	\label{tab:output2}
\end{table}

The CP violating measure $J_{\rm CP}$ is 
much smaller than the observed one.
As $g_q\,(q=d,\,u)$, are real, the CP violating phase is generated by 
${\rm Re} \,\tau$.
 This contribution is strongly suppressed for the same reason
it is suppressed  in the  quark flavour model 
with $A_4$ modular symmetry with real $g_d$ and $g_u$,
considered in the vicinity 
of the symmetric point $\tau=\omega$ in \cite{Petcov:2022fjf}.
The cause of the suppression is analysed in detail in Section 5 of 
ref. \cite{Petcov:2022fjf} and we will not repeat the analysis here. 
We only note that, as it follows from the discussion in \cite{Petcov:2022fjf}, 
the suppression of interest might not take place if some of the 
constants present in the down-type and up-type quark mass matrices 
$M_d$ and $M_u$ are complex and CP-violating and/or 
if $M_d$ and $M_u$ depend on two different CP-violating 
moduli $\tau_d$ and $\tau_u$, 
respectively, with $\tau_d \neq \tau_u$. In what follows we explore 
phenomenologically these possibilities.
%

%
\subsubsection{ Complex $g_d$ ($g_u$)}
%
%

 We assume first that $g_d$ is complex, while $g_u$ is real.
We present a sample set of the results of the fitting of the input data 
in Tables \ref{tab:input3} and  \ref{tab:output3}.

\begin{table}[h]
	\begin{center}
		\renewcommand{\arraystretch}{1.1}
		\begin{tabular}{|c|c|c|c|c|c|c|c|} \hline
			\rule[14pt]{0pt}{3pt}  
			$\tau$ & $\frac{\beta_d}{\alpha'_d}$ & $\frac{\gamma_d}{\alpha'_d}$ & $|g_d|$ &${\rm arg}\,[g_d]$ 
			& $\frac{\beta_u}{\alpha'_u}$ & $\frac{\gamma_u}{\alpha'_u}$ & $g_u$
			\\
			\hline
			\rule[14pt]{0pt}{3pt}  
			$-0.0920+i\, 2.394$   &$3.47$ & $1.27$  & $0.88$ &$161^\circ$ 
			 & $1.85$	& $1.15$&  $-10.54$\\ \hline
		\end{tabular}
	\end{center}
	\caption{
		Values of the constant parameters obtained in the fit of the 
		quark mass ratios, CKM mixing angles and of the CPV phase $\delta_{\rm CP}$ 
		with complex  $g_d$. See text for details.
	}
	\label{tab:input3}
\end{table}
\begin{table}[h]
	\small{
		\begin{center}
			\renewcommand{\arraystretch}{1.1}
			\begin{tabular}{|c|c|c|c|c|c|c|c|c|c|} \hline
				\rule[14pt]{0pt}{3pt}  
				& $\frac{m_s}{m_b}\hskip -1 mm\times\hskip -1 mm 10^2$ 
				& $\frac{m_d}{m_b}\hskip -1 mm\times\hskip -1 mm 10^4$& $\frac{m_c}{m_t}\hskip -1 mm\times\hskip -1 mm 10^3$&$\frac{m_u}{m_t}\hskip -1 mm\times\hskip -1 mm 10^6$&
				$|V_{us}|$ &$|V_{cb}|$ &$|V_{ub}|$&$|J_{\rm CP}|$& $\delta_{\rm CP}$
				\\
				\hline
				\rule[14pt]{0pt}{3pt}  
				Fit &$1.56$ & $ 8.87$
				& $2.65$ & $3.16 $&
				$0.2263$ & $0.0774$ & $0.00436$ &
				$6.7\hskip -1 mm\times\hskip -1 mm 10^{-5}$&$64.3^\circ$
				\\ \hline
				\rule[14pt]{0pt}{3pt}
				Exp	 &$1.82$ & $9.21$ 
				& $2.80$& $ 5.39$ &
				$0.2250$ & $0.0400$ & $0.00353$ &$2.8\hskip -1 mm\times\hskip -1 mm 10^{-5}$&$66.2^\circ$\\
				$1\,\sigma$	&$\pm 0.10$ &$\pm 1.02$ & $\pm 0.12$& $\pm 1.68$ &$ \pm 0.0007$ &
				$ \pm 0.0008$ & $ \pm 0.00013$ &$^{+0.14}_{-0.12}\hskip -1 mm\times \hskip -1 mm 10^{-5}$&$^{+ 3.4^\circ}_{-3.6^\circ}$\\ \hline  
			\end{tabular}
		\end{center}
		\caption{Results of the fit of the quark mass ratios, 
			CKM mixing angles,  $\delta_{\rm CP}$ and  $J_{\rm CP}$ with complex $\epsilon$ and 
			$g_d$. 'Exp' denotes the values of the observables at the GUT scale, 
			including $1\sigma$ errors, quoted in Eqs.\,\eqref{Datamass}, 
			\eqref{DataCKM-GUT}, \eqref{CKMphase} and \eqref{JCPGUT}  
			and obtained from the measured ones.  
		}
		\label{tab:output3}
	}
\end{table}
%

 The results collected in Table \ref{tab:output3}
  show that in this case it is possible to reproduce the
  observed value of the CPV phase $\delta_{\rm CP}$. However,
  owing to the fact that the $|V_{\rm cb}|$ and $|V_{\rm ub}|$
  elements of the CKM matrix are larger respectively by the
  factors of 1.9 and 1.2 than the central values of these observables
  at the GUT scale, 
  the $J_{\rm CP}$ factor is also larger by more than a factor of 2
  than the value that should be reproduced.

 We have investigated also the case of complex $g_d$ and real $g_u$.
We find that in this case it is impossible to reproduce the
  observed value of the CPV phase $\delta_{\rm CP}$ and of the 
$J_{\rm CP}$ factor: the values we obtain for these observables are 
too small. The problem with reproducing the value of $|V_{cb}|$ also 
persists.

%

%
\subsubsection{ Complex  $g_d$ and $g_u$}
%

We consider further the case of both $g_d$ and $g_u$ being complex. 
We present a sample set of the results of the fitting 
in Tables \ref{tab:input4} and  \ref{tab:output4}.
\begin{table}[h]
	\begin{center}
		\renewcommand{\arraystretch}{1.1}
		\begin{tabular}{|c|c|c|c|c|c|c|c|c|} \hline
			\rule[14pt]{0pt}{3pt}  
			$\tau$ & $\frac{\beta_d}{\alpha'_d}$ & $\frac{\gamma_d}{\alpha'_d}$ & $|g_d|$ &${\rm arg}\,[g_d]$ 
			& $\frac{\beta_u}{\alpha'_u}$ & $\frac{\gamma_u}{\alpha'_u}$ & $|g_u|$&${\rm arg}\,[g_u]$
			\\
			\hline
			\rule[14pt]{0pt}{3pt}  
			$-0.3862+i\, 2.4132$   &$3.60$ & $1.04$  & $0.86$ &$161.2^\circ$ 
			& $2.02$	& $1.34$&  $10.4$& $205.6^\circ$\\ \hline
		\end{tabular}
	\end{center}
	\caption{
		Values of the constant parameters obtained in the fit of the 
		quark mass ratios, CKM mixing angles and of the CPV phase $\delta_{\rm CP}$ 
		with complex  $g_d$ and $g_u$. See text for details.
	}
	\label{tab:input4}
\end{table}
\begin{table}[h]
	\small{
		\begin{center}
			\renewcommand{\arraystretch}{1.1}
			\begin{tabular}{|c|c|c|c|c|c|c|c|c|c|} \hline
				\rule[14pt]{0pt}{3pt}  
				& $\frac{m_s}{m_b}\hskip -1 mm\times\hskip -1 mm 10^2$ 
				& $\frac{m_d}{m_b}\hskip -1 mm\times\hskip -1 mm 10^4$& $\frac{m_c}{m_t}\hskip -1 mm\times\hskip -1 mm 10^3$&$\frac{m_u}{m_t}\hskip -1 mm\times\hskip -1 mm 10^6$&
				$|V_{us}|$ &$|V_{cb}|$ &$|V_{ub}|$&$|J_{\rm CP}|$& $\delta_{\rm CP}$
				\\
				\hline
				\rule[14pt]{0pt}{3pt}  
				Fit &$1.52$ & $ 6.59$
				& $2.81$ & $3.37 $&
				$0.2233$ & $0.076$ & $0.00400$ &
				$5.8\hskip -1 mm\times\hskip -1 mm 10^{-5}$&$60.6^\circ$
				\\ \hline
				\rule[14pt]{0pt}{3pt}
				Exp	 &$1.82$ & $9.21$ 
				& $2.80$& $ 5.39$ &
				$0.2250$ & $0.0400$ & $0.00353$ &$2.8\hskip -1 mm\times\hskip -1 mm 10^{-5}$&$66.2^\circ$\\
				$1\,\sigma$	&$\pm 0.10$ &$\pm 1.02$ & $\pm 0.12$& $\pm 1.68$ &$ \pm 0.0007$ &
				$ \pm 0.0008$ & $ \pm 0.00013$ &$^{+0.14}_{-0.12}\hskip -1 mm\times \hskip -1 mm 10^{-5}$&$^{+ 3.4^\circ}_{-3.6^\circ}$\\ \hline  
			\end{tabular}
		\end{center}
		\caption{Results of the fit of the quark mass ratios, 
			CKM mixing angles,  $\delta_{\rm CP}$ and  $J_{\rm CP}$ with complex $g_d$ and $g_u$. 'Exp' denotes the values of the observables at the GUT scale, 
			including $1\sigma$ errors, quoted in Eqs.\,\eqref{Datamass}, 
			\eqref{DataCKM-GUT}, \eqref{CKMphase} and \eqref{JCPGUT}  
			and obtained from the measured ones.  
		}
		\label{tab:output4}
	}
\end{table}
%

We find that, as in the case of complex $g_d$ and real 
$g_u$, it is possible to reproduce the
  observed value of the CPV phase $\delta_{\rm CP}$ 
but the value of the  $J_{\rm CP}$ factor is larger by more 
than a factor of 2 than the correct value.
(see Table \ref{tab:output4}). 
This can be traced again to the larger 
than the GUT scale  values of 
$|V_{\rm cb}|$ and $|V_{\rm ub}|$ obtained in the fit.
 We did not attempt to improve the description of 
$|V_{\rm cb}|$ and $|V_{\rm ub}|$ by varying $\tan \beta$ 
and taking into account the threshold correction effetcs 
in the RG evolution of these two observables.
{ Accounting for these effects warrants an independent 
comprehensive study.
}

\vspace{0.6cm}
 It follows from the results reported in the preceding subsections 
that  it is possible to reproduce the down-type and up-type 
quark mass hierarchies in the considered model with 
$|\beta_q|/|\alpha_q| \sim |\gamma_q|/|\alpha_q| \sim {\cal O}(1)\,(q=d,\,u)$, 
$|g_d| = |\alpha_d|/|\alpha'_d| \sim {\cal O}(1)$ and 
$|g_u|\sim {\cal O}(10)$  when  the down-type and up-type 
quark mass matrices depend on the same modulus $\tau$. 
However, reproducing the CP violation in the quark sector 
is problematic. In certain cases 
this is due to the fact that the values of $|V_{ub}|$ and/or $|V_{cb}|$ 
resulting from the fit are larger than the 
GUT scale values of these observables.

Thus, in the next subsection we investigate phenomenologically also 
the possibility of having two different 
moduli in the down-type and up-type quark mass matrices $\tau_d$  
and $\tau_u$ and thus two different small parameters 
$\epsilon_d$ and $\epsilon_u$, $\epsilon_d \neq \epsilon_u$ 
\footnote{ 
This possibility may be realised in modular invariant flavour models 
with multiple moduli,   
such as models based on simplectic modular symmetry 
\cite{Ding:2021iqp}, or, e.g., in models based on $A_4\times A_4$ 
modular symmetry. Constructing such a model is beyond the scope of 
the present study. 
}.
We consider  the cases of real CP-conserving $g_d$ and $g_u$ constants
as well as complex CP-violating $g_d$ and $g_u$ constants.
In both cases the two moduli $\tau_d$  
and $\tau_u$ have CP violating VEVs.

%
\subsection{Two moduli $\tau_d$ and $\tau_u$}
%
%

\subsubsection{Real $g_d$	and $g_u$}
We present a sample set of the results of the fitting in this case   
in Tables \ref{tab:input5} and  \ref{tab:output5}.

\begin{table}[H]
	{	
		\begin{center}
			\renewcommand{\arraystretch}{1.1}
			\begin{tabular}{|c|c|c|c|c|c|c|c|} \hline
				\rule[14pt]{0pt}{3pt}  
				$\tau_d$	&	$\tau_u$ & $\frac{\beta_d}{\alpha'_d}$ & $\frac{\gamma_d}{\alpha'_d}$ 
				& $g_d$ 
				& $\frac{\beta_u}{\alpha'_u}$ & $\frac{\gamma_u}{\alpha'_u}$ & $g_u$
				\\
				\hline
				\rule[14pt]{0pt}{3pt}  
				$0.4234+i\, 2.4672$ 	&	$-0.4528+i\, 3.5211$   &$3.57$ & $1.32$  & $-0.703$ &$1.70$
				& $0.82$&  $-0.774$\\ \hline
			\end{tabular}
		\end{center}
		\caption{
			Values of the constant parameters obtained in the fit of the 
			quark mass ratios, CKM mixing angles and of the CPV phase $\delta_{\rm CP}$ 
			in the case of  two moduli $\tau_d $  and $\tau_u$ .
		}
		\label{tab:input5}
	}
\end{table}
\begin{table}[H]
	\small{
		\begin{center}
			\renewcommand{\arraystretch}{1.1}
			\begin{tabular}{|c|c|c|c|c|c|c|c|c|c|} \hline
				\rule[14pt]{0pt}{3pt}  
				& $\frac{m_s}{m_b}\hskip -1 mm\times\hskip -1 mm 10^2$ 
				& $\frac{m_d}{m_b}\hskip -1 mm\times\hskip -1 mm 10^4$& $\frac{m_c}{m_t}\hskip -1 mm\times\hskip -1 mm 10^3$&$\frac{m_u}{m_t}\hskip -1 mm\times\hskip -1 mm 10^6$&
				$|V_{us}|$ &$|V_{cb}|$ &$|V_{ub}|$&$|J_{\rm CP}|$& $\delta_{\rm CP}$
				\\
				\hline
				\rule[14pt]{0pt}{3pt}  
				Fit &$1.53$ & $ 7.90$
				& $2.51$ & $10.27$&
				$0.2238$ & $0.0467$ & $0.00342$ &
				$7.7\hskip -1 mm\times\hskip -1 mm 10^{-6}$&$12.8^\circ$
				\\ \hline
				\rule[14pt]{0pt}{3pt}
				Exp	 &$1.82$ & $9.21$ 
				& $2.80$& $ 5.39$ &
				$0.2250$ & $0.0400$ & $0.00353$ &$2.8\hskip -1 mm\times\hskip -1 mm 10^{-5}$&$66.2^\circ$\\
				$1\,\sigma$	&$\pm 0.10$ &$\pm 1.02$ & $\pm 0.12$& $\pm 1.68$ &$ \pm 0.0007$ &
				$ \pm 0.0008$ & $ \pm 0.00013$ &$^{+0.14}_{-0.12}\hskip -1 mm\times \hskip -1 mm 10^{-5}$&
				$^{+ 3.4^\circ}_{-3.6^\circ}$\\ \hline  
			\end{tabular}
		\end{center}
		\caption{Results of the fits of the quark mass ratios, 
			CKM mixing angles, $J_{\rm CP}$ and $\delta_{\rm CP}$ in the vicinity of two different 
			moduli in the down-quark and up-quark sectors, $\tau_d$ and $\tau_u$, 
			$\tau_d \neq \tau_u$. 'Exp' denotes the  values of the observables 
			at the GUT scale, including $1\sigma$ error, quoted in Eqs.\,\eqref{Datamass}, 
			\eqref{DataCKM-GUT}, \eqref{CKMphase} and \eqref{JCPGUT} 
			and obtained from the measured ones.
		}
		\label{tab:output5}
	}
\end{table}

 It follows from Table \ref{tab:output5} 
that the CP violation in the quark sector cannot be reproduced:
the value of the CPV phase $\delta_{\rm CP}$ is approximately 
by a factor of 5 smaller than the observed value at the GUT scale. 
Correspondingly, the $J_{\rm CP}$ factor is also smaller than 
the observed value.

%
\subsubsection{Complex  $g_d$ and $g_u$ }
%
%
We consider next the case of both 
constants $g_d$ and $g_u$ being complex. 
A sample set of the results of the fitting in this case 
is presented in Tables \ref{tab:input6} and  \ref{tab:output6}.


 As follows from Tables\,\ref{tab:input6} and \ref{tab:output6},
all quark observables can be successfully reproduced 
with all constant being in magnitude of the same order.
The magnitude of the measure of goodness of the fitting $N\sigma$,
which is defined  in Appendix \ref{fit}, is 
$N\sigma=2.8$. The number of parameters is rather large -- 
this successful model has altogether 14 real parameters 
(10 real constants and 4 phases). 
In the next subsection we consider alternative models 
with a common modulus $\tau$ in the down-type and up-type 
quark mass matrices, with smaller number of parameters.

\begin{table}[H]
\small{	
		\begin{center}
			\renewcommand{\arraystretch}{1.1}
			\begin{tabular}{|c|c|c|c|c|c|c|c|c|c|} \hline
				\rule[14pt]{0pt}{3pt}  
				$\tau_d$	&	$\tau_u$ & $\frac{\beta_d}{\alpha'_d}$ & $\frac{\gamma_d}{\alpha'_d}$ & $|g_d|$ &${\rm arg}\,[g_d]$ 
				& $\frac{\beta_u}{\alpha'_u}$ & $\frac{\gamma_u}{\alpha'_u}$ & $|g_u|$&${\rm arg}\,[g_u]$
				\\
				\hline
				\rule[14pt]{0pt}{3pt}  
				$-0.3087+i\, 2.4012$ 	&	$0.1231+i\, 3.8110$   &$4.60$ & $1.33$  & $0.80$ &$198.5^\circ$  & $3.00$
				& $2.10$&  $1.22$& $218.3^\circ$\\ \hline
			\end{tabular}
		\end{center}
		\caption{
			Values of the constant parameters obtained in the fit of the 
			quark mass ratios, CKM mixing angles and of the CPV phase $\delta_{\rm CP}$ 
			in the case of  two moduli $\tau_d$ and $\tau_u$.
		}
\label{tab:input6}
}
\end{table}

\begin{table}[H]
\small{
		\begin{center}
			\renewcommand{\arraystretch}{1.1}
			\begin{tabular}{|c|c|c|c|c|c|c|c|c|c|} \hline
				\rule[14pt]{0pt}{3pt}  
				& $\frac{m_s}{m_b}\hskip -1 mm\times\hskip -1 mm 10^2$ 
				& $\frac{m_d}{m_b}\hskip -1 mm\times\hskip -1 mm 10^4$& $\frac{m_c}{m_t}\hskip -1 mm\times\hskip -1 mm 10^3$&$\frac{m_u}{m_t}\hskip -1 mm\times\hskip -1 mm 10^6$&
				$|V_{us}|$ &$|V_{cb}|$ &$|V_{ub}|$&$|J_{\rm CP}|$& $\delta_{\rm CP}$
				\\
				\hline
				\rule[14pt]{0pt}{3pt}  
				Fit &$1.76$ & $ 7.74$
				& $2.84$ & $5.61 $&
				$0.2251$ & $0.0414$ & $0.00337$ &
				$2.83\hskip -1 mm\times\hskip -1 mm 10^{-5}$&$67.7^\circ$
				\\ \hline
				\rule[14pt]{0pt}{3pt}
				Exp	 &$1.82$ & $7.93$ 
				& $2.82$& $ 6.55$ &
				$0.2250$ & $0.0400$ & $0.00353$ &$2.8\hskip -1 mm\times\hskip -1 mm 10^{-5}$&$66.2^\circ$\\
				$1\,\sigma$	&$\pm 0.10$ &$\pm 1.02$ & $\pm 0.12$& $\pm 1.68$ &$ \pm 0.0007$ &
				$ \pm 0.0008$ & $ \pm 0.00013$ &$^{+0.14}_{-0.12}\hskip -1 mm\times \hskip -1 mm 10^{-5}$&
				$^{+ 3.4^\circ}_{-3.6^\circ}$\\ \hline  
			\end{tabular}
		\end{center}
		\caption{Results of the fits of the quark mass ratios, 
			CKM mixing angles, $J_{\rm CP}$ and $\delta_{\rm CP}$ in the vicinity of two different 
			moduli in the down-quark and up-quark sectors, $\tau_d$ and $\tau_u$, 
			$\tau_d \neq \tau_u$. 'Exp' denotes the  values of the observables 
			at the GUT scale, including $1\sigma$ error, quoted in Eqs.\,\eqref{Datamass}, 
			\eqref{DataCKM-GUT}, \eqref{CKMphase} and \eqref{JCPGUT} 
			and obtained from the measured ones.
		}
		\label{tab:output6}
}
\end{table}

%
\subsection{Alternative model with a common modulus $\tau$}	
\label{sub-weight8}

Finally, we discuss  an alternative model. In this model we 
introduce  weight 8 modular forms  
in addition to the weights 4 and 6 ones in an attempt  to get 
a correct description  of the observed three CKM mixing angles 
and CP violating phase with one modulus $\tau$.
The model is obtained from the considered one by 
replacing in Table \ref{tab:model} the weights  $(6,\,2\,,0)$ of
the  right-handed  quarks $(u^c,c^c,t^c)$.
 The weights are same ones 
$(4,\,2\,,0)$	for right-handed quarks $(d^c,s^c,b^c)$.
These assigments  are summarized in Table \ref{tab:model-8}.
\begin{table}[H]
	\begin{center}
		\renewcommand{\arraystretch}{1.1}
		\begin{tabular}{|c|c|c|c|c|} \hline
			& $Q$ &
			 $(d^c\,,s^c\,,b^c)\ , (u^c\,,c^c\,,t^c)$ &  $H_u$ & $H_d$ \\ \hline
			$SU(2)$ & 2 & 1  & 2 & 2 \\
			$A_4$ & 3 & $(1',\,1',1')$\quad\ $(1',\,1',1')$ & $1$ & $1$ \\
			$k$ & 2 & $(4,\,2,\,0)$\qquad  $(6,\,2,\,0)$& 0 & 0 \\ \hline
		\end{tabular}
	\end{center}
	\caption{Assignments of $A_4$ representations and weights in our model.}
	\label{tab:model-8}
\end{table}
%
Then, the quark mass matrices are easily obtained as:
\begin{align}
&  { M_d} =v_d
\begin{pmatrix}
\hat\alpha'_d  & 0 & 0  \\
0 & \hat\beta_d & 0 \\
0& 0 & \hat\gamma_d \\
\end{pmatrix}
\begin{pmatrix}
\tilde Y_3^{(6)} &  \tilde Y_2^{(6)} &  \tilde Y_1^{(6)} \\
\tilde Y_3^{(4)} &  \tilde Y_2^{(4)} &  \tilde Y_1^{(4)} \\
Y^{(2)}_3 & Y^{(2)}_2 &Y^{(2)}_1 
\end{pmatrix},\ \ 
{ M_u} =v_u
\begin{pmatrix}
\hat\alpha'_u  & 0  & 0  \\
0 & \hat\beta_u & 0 \\
0& 0 & \hat\gamma_u \\
\end{pmatrix}
\begin{pmatrix}
\tilde Y_3^{(8)} &  \tilde Y_2^{(8)} &  \tilde Y_1^{(8)} \\
\tilde Y_3^{(4)} &  \tilde Y_2^{(4)} &  \tilde Y_1^{(4)} \\
Y^{(2)}_3 & Y^{(2)}_2 &Y^{(2)}_1 \\
\end{pmatrix},
\label{Massmatrix-8}
\end{align}
%
where
\begin{align}
\tilde Y_i^{(6)}=  g_{d} Y_i^{(6)} +   Y_i^{'(6)}\, , \qquad 
\tilde Y_i^{(8)}=  f_{u} Y_i^{(8)} + Y_i^{'(8)}\, ,  \qquad 
g_{d}\equiv\alpha_d/\alpha'_d\, \qquad
f_{u} \equiv \alpha_u/\alpha'_u\,.
\end{align}
%
In order to get the canonical form of the  kinetic term, 
as discussed in Eq.\,\eqref{canonical},
the couplings are  shifted  by  overall normalizations as follows:
\begin{eqnarray}
&&\alpha_u \rightarrow \hat\alpha_u= \alpha_u\, \sqrt{(2 {\rm Im} \tau)^{8} }
=\alpha_u\,(2 {\rm Im} \tau)^{4},\, \quad
\alpha'_u \rightarrow \hat\alpha'_u= \alpha'_u\, \sqrt{(2 {\rm Im} \tau)^{8} }
=\alpha'_u\,(2 {\rm Im} \tau)^{4}\, ,\nonumber\\
&&\beta_u  \rightarrow \hat\beta_u = \beta_u  \, \sqrt{(2 {\rm Im} \tau)^{4} }
= \beta_u  \, (2 {\rm Im} \tau)^2\, , \quad\
\gamma_u  \rightarrow \hat\gamma_u = \gamma_u \sqrt{(2 {\rm Im} \tau)^{2} }
=\gamma_u \, (2 {\rm Im} \tau)\, , \nonumber\\
&&\alpha_d \rightarrow \hat\alpha_d= \alpha_d\, \sqrt{(2 {\rm Im} \tau)^{6} }
=\alpha_d\,(2 {\rm Im} \tau)^{3},\, \quad\
\alpha'_d \rightarrow \hat\alpha'_d= \alpha'_d\, \sqrt{(2 {\rm Im} \tau)^{6} }
=\alpha'_d\,(2 {\rm Im} \tau)^{3}\, ,\nonumber\\
&&\beta_d \rightarrow \hat\beta_d = \beta_d  \, \sqrt{(2 {\rm Im} \tau)^{4} }
= \beta_d  \, (2 {\rm Im} \tau)^2\, , \quad\ \
\gamma_d  \rightarrow \hat\gamma_d = \gamma_d \sqrt{(2 {\rm Im} \tau)^{2} }
=\gamma_d \, (2 {\rm Im} \tau)\, . 
\label{shift2}
\end{eqnarray}
%
In the considered model we have two parameters $g_d$ and   $f_u$ 
in addition to $\alpha'_q$, $\beta_q$ and $\gamma_q$.
The elements of the matrix ${ M^\dagger_d M_d}$ are given in 
Eq.\,(\ref{masscomponent}). We give below the elements of 
${ (M^\dagger_u M_u)_{ij}} \equiv {\cal M}^2_u[i,j]$ in units of $v^2_u$ 
to leading order in $\epsilon$:
\begin{align}
&{\cal M}_u^{2} [1,1]=
324\,\epsilon^4 \left [|\hat\gamma_u|^2+
9|\hat\beta_u|^2 + |\hat\alpha'_u|^2 |3f_u - 4|^2\right ]\,,\nonumber\\
&{\cal M}_u^{2} [2,2]=
36\,\epsilon^2 \left [|\hat\gamma_u|^2+
|\hat\beta_u|^2 + |\hat\alpha'_u|^2|f_u - 2|^2\right ]\,,\nonumber\\
&{\cal M}_u^{2} [3,3]=
|\hat\gamma_u|^2+
|\hat\beta_u|^2 + |\hat\alpha'_u|^2|f_u|^2\,,\nonumber\\
&{\cal M}_u^{2} [1,2]= 108\,\epsilon^3\,p^* \left [|\hat\gamma_u|^2+
3|\hat\beta_u|^2 + |\hat\alpha'_u|^2(f_u - 2)(3f^*_u-4)\right ]\,,\nonumber\\
&{\cal M}_u^{2} [1,3]= 18\,\epsilon^2 \, p^{*2} \left [-\,|\hat\gamma_u|^2 + 
3|\hat\beta_u|^2 + |\hat\alpha'_u|^2\,f_u(3f^*_u-4)\right ]\,,\nonumber\\
&{\cal M}_u^{2} [2,3]= 6\,\epsilon\, p^{*} \left [-\,|\hat\gamma_u|^2 + 
|\hat\beta_u|^2 + |\hat\alpha'_u|^2\, f_u(f^*_u-2)\right ]\,,\nonumber\\
&{\cal M}_u^{2} [2,1]={\cal M}_u^{2} [1,2]^*\,,\qquad
{\cal M}_u^{2} [3,1]={\cal M}_u^{2} [1,3]^*\,,\qquad
{\cal M}_u^{2} [3,2]={\cal M}_u^{2} [2,3]^*\,.
\label{MassCompII}
\end{align}
%
Using these results we obtain to leading order in $\epsilon$: 
\begin{align}
m_t^2\simeq {\rm Tr\,({\cal M}^2_u)}\simeq 
v^2_u (|\hat\alpha'_u|^2\,|f_u|^2 + |\hat\beta_u|^2 + 
|\hat\gamma_u|^2)\,,
\end{align}
and, from the the determinant of the submatric of the 2-3 sector, 
\begin{align}
m_c^2 m_t^2 \simeq (12)^2 \,v^4_u\,\epsilon^2
(|\hat\alpha'_u|^2\,|\hat\beta_u|^2 + 
(1 - {\rm Re}(f_u))^2\,|\hat\alpha'_u|^2\,|\hat\gamma_u|^2 + 
|\hat\beta_u|^2\,|\hat\gamma_u|^2) \,.
\end{align}
%
We find also that the 
determinant of $ {\cal M}^2_u$ vanishes:
\begin{align}
{\rm Det}\, [{ \cal M}^2_u] = 0\,.
\end{align}
%
As shown in Appendix \ref{Det=0}, 
the vanishing of ${\rm Det}\, [{ \cal M}^2_u]$
is exact: it follows from the expression of $ M_u$ 
in Eq.\,\eqref{Massmatrix-8} and is a consequence of 
the properties of the modular forms given in Eq.\,(\ref{condition}) 
and the relation  ${\bf Y^{(\rm 8)}_3} = (Y_1^2+2 Y_2 Y_3){\bf Y^{(\rm 4)}_3}$.
As a consequence, $ M_u$ is a rank two matrix and   
the lightest quark, $u$-quark, is massless.
 According to \cite{ParticleDataGroup:2022pth}, given 
 the results of various estimates of the value of $m_u$ 
 (lattice calculations, chiral perturbation theory, etc.),
 $m_u = 0$
 seems very unlikely and  strongly disfavored.
 However, 
there exist mechanisms which  
generate the requisite tiny $u$-quark mass without 
affecting the other predictions of the model.
 
 The $u$-quark mass $m_u\neq 0$ could be  generated by supersymmetry 
breaking \cite{Feruglio:2021dte}.
 If supersymmetry is broken by some F-term, the Yukawa couplings are corrected 
 by terms of the order of  $F/\Lambda^2$, where $F$ is the supersymmetry 
breaking expectation value with dimension of  mass square and 
$\Lambda$ is the SUSY breaking messenger scale \cite{Criado:2018thu,Feruglio:2021dte}
\\

As a second possibility of generating  $m_u\neq 0$, one can consider  
the dimension six operators \cite{Feruglio:2021dte}:
\begin{align}
(u^c Q H_u)(H_u H_d) \,,\qquad (c^c Q H_u)(H_u H_d) \,,
\qquad (t^c Q H_u)(H_u H_d) \,,
\label{new-operator}
\end{align}
%
whose Wilson coefficient should be a modular form of the appropriate weight. 
In order for this mechanism to work, the weight assignments in Table 
\ref{tab:model-8} should be modified. The following conditions 
have to be fulfilled:
\begin{align}
k_Q=2-k_{Hd}\,,\quad k_{u^c}=6+k_{Hd}-k_{Hu}\,,
\quad k_{c^c}={2}+k_{Hd}-k_{Hu}\,
\quad k_{t^c}= k_{Hd}-k_{Hu}\,,
\label{Condition-1}
\end{align}
%
with the additional constraint 
\begin{align}
 k_{Hd}+k_{Hu}\not = 0\,,
 \label{Condition-2}
\end{align}
%
where $k_{Hd}$ and $k_{Hu}$ denote weights of $H_d$ and $H_u$, respectively.
The  conditions in Eq.\,\eqref{Condition-1} ensure that the superpotentials 
discussed in the previous subsections have weight zero,
and the one in Eq.\,\eqref{Condition-2} implies that the operator
of Eq.\,\eqref{new-operator} has different weight from
the corresponding renormalisable Yukawa term, so that it couples to an 
independent modular form multiplet with weight $8+k_{Hd}+k_{Hu}$,
$4+k_{Hd}+k_{Hu}$, $2+k_{Hd}+k_{Hu}$.
These terms make  the resulting up-type quark mass matrix of rank three.
Such a mechanism generates tiny quark mass $m_u\sim v_dv_u^2/\Lambda^2$,
where $\Lambda$ is the scale at which the operator in 
Eq.\,\eqref{new-operator} is generated.

Adding the tiny corrections due to the SUSY breaking   or
the  dimension six operator discussed above,
the up-type quark mass matrix $M_u$ is modified as follows: 
\begin{align}
M_u =v_u
\begin{pmatrix}
\hat\alpha'_u  & 0  & 0  \\
0 & \hat\beta_u & 0 \\
0& 0 & \hat\gamma_u \\
\end{pmatrix}
\begin{pmatrix}
\tilde Y_3^{(8)}(1+C_{u1}) &  \tilde Y_2^{(8)} &  \tilde Y_1^{(8)} \\
\tilde Y_3^{(4)}(1+C_{u2})  &  \tilde Y_2^{(4)} &  \tilde Y_1^{(4)} \\
Y^{(2)}_3(1+C_{u3})  & Y^{(2)}_2 &Y^{(2)}_1 \\
\end{pmatrix},
\label{Massmatrix-breaking}
\end{align}
%
where the  mass correction terms appear in the first column.
Their contribution
are negligibly small  in the second and third columns 
for $\tau$ close to  $\tau=i\infty$ (e.g., ${\rm Im}\tau \sim 2.5$). 
The contribution to the down-type quark mass matrix is negligible as well. 
Unless $C_{u1}=C_{u2}=C_{u3}$, the determinant of ${\cal M}_u^2$ is non-vanishing 
in this case:
\begin{align}
{\rm Det}[{\cal M}_u^2] =(6)^6\hat\alpha'^2_q\hat\beta_q^2 \hat\gamma_q^2 v_q^6\, \epsilon^6 \, |C_{u}|^2 \,,
\end{align}
%
where
\begin{align}
 C_{u}
=3 f_u\, (C_{u1}-C_{u2})+(-4 C_{u1}+3 C_{u2}+C_{u3})\,.
\end{align}
%
Thus, ${\rm Det}\,[{\cal M}_q^2]$ depends on $|C_{u}|^2$.
Indeed, as shown numerically below, the observed $u$-quark mass is reproduced 
for
$|C_{u}|\sim 0.1$ and $|f_u|={\cal O}(1)$.
The contributions of $|C_{ui}| \sim 0.1\,(i=1,2,3)$ 
to 
the quark mixing angles and the CPV phase are negligible 
because they are much smaller than the magnitudes of
modular forms $\tilde Y_3^{(k)}$.


 In the considered model 
the up-type quark mass hierarchy differs from that discussed
	after Eq.\,\eqref{subdet} in subsection \ref{Quark642}. 
	Suppose that $|\alpha'_u|$, $|\beta_u|$ and $|\gamma_u|$
	are of the same order.
Bringing the kinetic terms to their canonical forms 
leads to normalisation factors of $\alpha'_u$, $\beta_u$ and $\gamma_u$ 
which differ from those in Eq.\,\eqref{shift}
since the weights of the right-handed up-type quarks are different 
in the considered alternative model
($k_{u^c}=6$, $k_{c^c}=2$, $k_{y^c}=0$). 
After taking into account that 
due to the renormalisation factors 
$|\hat\alpha'_u| \gg |\hat\beta_u| \gg |\hat\gamma_u|$ 
and assuming for simplicity that $f_u$ is real and 
$f_u \gtrsim 1$ we have:
	\begin{align}
	m_{t} \simeq \hat\alpha'_u\, f_u=\alpha'_q I_\tau^4\, f_u, \ \
	m_{c} \simeq \hat\beta_u\frac{1}{f_u} (12\,\epsilon)=\beta_q I_\tau^2 \frac{1}{f_u}(12\,\epsilon),\ \
	m_{u} \simeq 6^3\hat\gamma_u\epsilon^2 |C_u|
	=6^3\gamma_u I_\tau \epsilon^2 |C_u|,
	\end{align}
%
where as before $I_\tau=2{\rm Im}\,\tau$.
Thus, in the case of  $|\alpha'_u|\sim |\beta_u| \sim |\gamma_u|$
for the mass ratios we get:
	\begin{align}
	m_u : m_{c}: m_{t}&\simeq
	I_\tau^4 f_u:12 I_\tau^2 \frac{1}{f_u}\epsilon:   6^3I_\tau\epsilon^2 |C_u|
	\nonumber\\
	&= \left [1:\left (\frac{12\epsilon}{I_\tau f_u} 
\frac{1}{I_\tau f_u}\right ): 
	\frac32\left(\frac{12\epsilon}{I_\tau f_u}
	\frac{1}{I_\tau f_u}\right)^2 f_u^3 I_\tau|C_u|\right]\,I_\tau^4 f_u \,.
	\label{up-Hiererchy}
	\end{align}
%
The observed up-type quark mass hierarchies are reproduced
for ${\rm Im}\,\tau \sim 2.5$ and small $|C_u|$ with 
$f_u={\cal O}(1)$. For example, we get $m_u/m_t= 5\times 10^{-6}$
if  ${\rm Im}\,\tau=2.5$ and $|C_{u}/f_u|=0.1$ in Eq.\,\eqref{up-Hiererchy}.
The ratios of the up-type quark masses should be compared with
the down-type quark mass ratios:
\begin{align}
1: \frac{m_{s}}{m_{b}}:\frac{m_{d}}{m_{b}} \simeq 
1: \left(\frac{12\epsilon}{I_\tau g_d}\right): 
\left(\frac{12\epsilon}{I_\tau g_d}\right)^2\,.
\end{align}
%
We note that the relatively large value of ${\rm Im}\,\tau$ 
together with  $|C_{u}|\sim 0.1$ 
gives the requisite stronger hierarchies of up-type quarks masses 
as compared to the down-type quark mass hierarchies 
even if  $f_u={\cal O}(1)$. 

In what follows we show  samples of results of the numerical 
analysis of the model.
We take $C_{u1}$ to be real  and neglect $C_{u2}$ and $C_{u3}$ for simplicity.

%
 \subsubsection{ Real $g_d$ and $f_u$}
 %

 In the case of real  $g_d$ and $f_u$ the model 
contains 11 real parameters.
A tentative  sample set of  fitting  of the  values of the observables 
at the GUT scale, given in Section \ref{sec:inputs}, 
 is shown in Tables \ref{tab:input-alt1} and  \ref{tab:output-alt1}.

 \begin{table}[h]
 	\begin{center}
 		\renewcommand{\arraystretch}{1.1}
 		\begin{tabular}{|c|c|c|c|c|c|c|c|} \hline
 			\rule[14pt]{0pt}{3pt}  
 			$\tau$ & $\frac{\beta_d}{\alpha'_d}$ & $\frac{\gamma_d}{\alpha'_d}$ & $g_d$ 
 			& $\frac{\beta_u}{\alpha'_u}$ & $\frac{\gamma_u}{\alpha'_u}$ & $f_u$
 			&$C_{u1}$\\
 			\hline
 			\rule[14pt]{0pt}{3pt}  
 			$-0.3416+i\,2.3818$   &$3.89$ & $1.12$  & $-0.66$  & $1.92$
 			& $3.45$&  $-1.78$&$-0.069$\\ \hline
 		\end{tabular}
 	\end{center}
 	\caption{ Values of the constant parameters obtained in the fit of the 
 		quark mass ratios and of 
 		CKM mixing angles. 
 		See text for details.
 	\label{tab:input-alt1}
 	}
 \end{table}
 
 \begin{table}[h]
 	\begin{center}
 		\renewcommand{\arraystretch}{1.1}
 		\begin{tabular}{|c|c|c|c|c|c|c|c|c|} \hline
 			\rule[14pt]{0pt}{3pt}  
 			& $\frac{m_s}{m_b}\hskip -1 mm\times\hskip -1 mm 10^2$ 
 			& $\frac{m_d}{m_b}\hskip -1 mm\times\hskip -1 mm 10^4$& $\frac{m_c}{m_t}\hskip -1 mm\times\hskip -1 mm 10^3$&$\frac{m_u}{m_t}\hskip -1 mm\times\hskip -1 mm 10^6$&
 			$|V_{us}|$ &$|V_{cb}|$ &$|V_{ub}|$&$J_{\rm CP}$
 			\\
 			\hline
 			\rule[14pt]{0pt}{3pt}  
 			Fit &$2.09$ & $ 8.90$
 			& $3.14$ & $6.60 $&
 			$0.2267$ & $0.0296$ & $0.00382$ &
 			$7.0\hskip -1 mm\times\hskip -1 mm 10^{-9}$
 			\\ \hline
 			\rule[14pt]{0pt}{3pt}
 			Exp	 &$1.82$ & $9.21$ 
 			& $2.80$& $ 5.39$ &
 			$0.2250$ & $0.0400$ & $0.00353$ &$2.8\hskip -1 mm\times\hskip -1 mm 10^{-5}$\\
 			$1\,\sigma$	&$\pm 0.10$ &$\pm 1.02$ & $\pm 0.12$& $\pm 1.68$ &$ \pm 0.0007$ &
 			$ \pm 0.0008$ & $ \pm 0.00013$ &$^{+0.14}_{-0.12}\hskip -1 mm\times \hskip -1 mm 10^{-5}$\\ \hline  
 		\end{tabular}
 	\end{center}
 	\caption{Results of the fit of the quark mass ratios and 
 		CKM mixing angles.  $J_{\rm CP}$ factor is output.
 		'Exp' denotes the respective values at the GUT scale, 
 		including $1\sigma$ errors.
 		\label{tab:output-alt1}
 	}
 \end{table}
%

 Since $g_d$ and $f_u$ are real and thus CP-conserving, 
the value of the CP violating measure $J_{\rm CP}$ is much smaller than the 
value of  observed one at the GUT scale. The reason is 
practically identical to that discussed in  \cite{Petcov:2022fjf}.
By introducing an additional source of the CP violation in the form of 
complex $g_d$ or/and $f_u$, the quality of the 
fit is improved considerably, as is shown below.
%
 \subsubsection{ Real $g_d$ and complex $f_u$}
%

 This is the case with minimum number of parameters 
in which we can possibly have sufficiently large 
violation of the CP symmetry in the quark sector.
The CP violation is generated by the 
complex parameter $f_u$ and the modulus $\tau$.
The total number of parameters is 12 (10 real constants and 2 phases).
The numerical results of the fitting are presented
in Tables \ref{tab:input-alt2} and \ref{tab:output-alt2}.
\begin{table}[h]
			\begin{center}
				\renewcommand{\arraystretch}{1.1}
				\begin{tabular}{|c|c|c|c|c|c|c|c|c|} \hline
					\rule[14pt]{0pt}{3pt}  
					$\tau$ & $\frac{\beta_d}{\alpha'_d}$ & $\frac{\gamma_d}{\alpha'_d}$ & $g_d$ 
					& $\frac{\beta_u}{\alpha'_u}$ & $\frac{\gamma_u}{\alpha'_u}$ & $|f_u|$&${\rm arg}\,[f_u]$	& $C_{u1}$
					\\
					\hline
					\rule[14pt]{0pt}{3pt}  
					$-0.3952+i\, 2.4039$   &$3.82$ & $1.17$  & $-0.677$ 
					& $1.72$	& $3.21$&  $1.68$& $127.3^\circ$& -0.07147\\ \hline
				\end{tabular}
			\end{center}
			\caption{
				Values of the constant parameters obtained in the fit of the 
				quark mass ratios, CKM mixing angles and of the CPV phase $\delta_{\rm CP}$ 
				with real $C_{u1}$, real $g_d$ and complex $f_u$ . See text for details.}
	\label{tab:input-alt2}
\end{table}
\vskip -0.5 cm
\begin{table}[H]
	\small{ 
			\begin{center}
				\renewcommand{\arraystretch}{1.1}
				\begin{tabular}{|c|c|c|c|c|c|c|c|c|c|} \hline
					\rule[14pt]{0pt}{3pt}  
					& $\frac{m_s}{m_b}\hskip -1 mm\times\hskip -1 mm 10^2$ 
					& $\frac{m_d}{m_b}\hskip -1 mm\times\hskip -1 mm 10^4$& $\frac{m_c}{m_t}\hskip -1 mm\times\hskip -1 mm 10^3$&$\frac{m_u}{m_t}\hskip -1 mm\times\hskip -1 mm 10^6$&
					$|V_{us}|$ &$|V_{cb}|$ &$|V_{ub}|$&$|J_{\rm CP}|$& $\delta_{\rm CP}$
					\\
					\hline
					\rule[14pt]{0pt}{3pt}  
					Fit &$1.89$ & $ 8.78$
					& $2.81$ & $5.52 $&
					$0.2251$ & $0.0390$ & $0.00364$ &
					$2.94\hskip -1 mm\times\hskip -1 mm 10^{-5}$&$70.7^\circ$
					\\ \hline
					\rule[14pt]{0pt}{3pt}
					Exp	 &$1.82$ & $9.21$ 
					& $2.80$& $ 5.39$ &
					$0.2250$ & $0.0400$ & $0.00353$ &$2.8\hskip -1 mm\times\hskip -1 mm 10^{-5}$&$66.2^\circ$\\
					$1\,\sigma$	&$\pm 0.10$ &$\pm 1.02$ & $\pm 0.12$& $\pm 1.68$ &$ \pm 0.0007$ &
					$ \pm 0.0008$ & $ \pm 0.00013$ &$^{+0.14}_{-0.12}\hskip -1 mm\times \hskip -1 mm 10^{-5}$&$^{+ 3.4^\circ}_{-3.6^\circ}$\\ \hline  
				\end{tabular}
\end{center}
}
\caption{Results of the fit of the quark mass ratios, 
CKM mixing angles,  $\delta_{\rm CP}$ and  $J_{\rm CP}$ with real $g_d$ and 
complex $f_u$. 'Exp' denotes the values of the observables at the GUT scale, 
including $1\sigma$ errors, quoted in Eqs.\,\eqref{Datamass}, 
\eqref{DataCKM-GUT}, \eqref{CKMphase} and \eqref{JCPGUT}  
and obtained from the measured ones. 
}
\label{tab:output-alt2}
\end{table}
%

As follows from Tables \ref{tab:input-alt2} and \ref{tab:output-alt2}
	all quark observables can be successfully reproduced 
	with all constant being in magnitude of the same order.
	The magnitude of the measure of goodness of the fit $N\sigma$
	is $N\sigma = 2.0$.
We emphasise that  both the mass hierarchies of down-type and up-type 
quarks are reproduced by the common modulus $\tau$ with 
one complex parameter  $f_u$, which is of order one in magnitude.
%
%

%
\subsubsection{ Complex  $g_d$ and $f_u$}
%
%
The case of  the complex $g_d$ with  real $f_u$ does not give us a 
fitting result with $N\sigma$  smaller than $N\sigma=5$, 
so, we omit the discussion of this case.
 
In the case with   both complex  $g_d$   and  $f_u$ 
the number of parameters is  altogether 13.  
The sources of the CP violation in this case are  
the phases of the modulus $\tau$ and of the constants 
$g_d$ and $f_u$. The numerical results are  
presented in Tables \ref{tab:input-alt4} and  \ref{tab:output-alt4}. 

\begin{table}[H]
	{ 
			\begin{center}
				\renewcommand{\arraystretch}{1.1}
				\begin{tabular}{|c|c|c|c|c|c|c|c|c|c|} \hline
					\rule[14pt]{0pt}{3pt}  
					$\tau$ & $\frac{\beta_d}{\alpha'_d}$ & $\frac{\gamma_d}{\alpha'_d}$ & $|g_d|$ &${\rm arg}\,[g_d]$ 
					& $\frac{\beta_u}{\alpha'_u}$ & $\frac{\gamma_u}{\alpha'_u}$ & $|f_u|$&${\rm arg}\,[f_u]$	& $C_{u1}$
\\
\hline
					\rule[14pt]{0pt}{3pt}  
					$-0.3629+i\, 2.4090$   &$3.82$ & $1.06$  & $0.68$ &$177.7^\circ$ 
					& $1.60$	& $3.28$&  $1.61$& $122.5^\circ$& -0.0707
\\ 
\hline
\end{tabular}
\end{center}
\caption{
				Values of the constant parameters obtained in the fit of the 
				quark mass ratios, CKM mixing angles and of the CPV phase $\delta_{\rm CP}$ 
				with real $C_{u1}$, and  complex  $g_d$ and  $f_u$. See text for details.}
\label{tab:input-alt4}
}
\end{table}
\vskip -0.5 cm
\begin{table}[H]
		\small{	\begin{center}
				\renewcommand{\arraystretch}{1.1}
				\begin{tabular}{|c|c|c|c|c|c|c|c|c|c|} \hline
					\rule[14pt]{0pt}{3pt}  
					& $\frac{m_s}{m_b}\hskip -1 mm\times\hskip -1 mm 10^2$ 
					& $\frac{m_d}{m_b}\hskip -1 mm\times\hskip -1 mm 10^4$& $\frac{m_c}{m_t}\hskip -1 mm\times\hskip -1 mm 10^3$&$\frac{m_u}{m_t}\hskip -1 mm\times\hskip -1 mm 10^6$&
					$|V_{us}|$ &$|V_{cb}|$ &$|V_{ub}|$&$|J_{\rm CP}|$& $\delta_{\rm CP}$
					\\
					\hline
					\rule[14pt]{0pt}{3pt}  
					Fit &$1.86$ & $ 7.77$
					& $2.86$ & $5.29 $&
					$0.2251$ & $0.0398$ & $0.00357$ &
					$2.87\hskip -1 mm\times\hskip -1 mm 10^{-5}$&$67.1^\circ$
					\\ \hline
					\rule[14pt]{0pt}{3pt}
					Exp	 &$1.82$ & $9.21$ 
					& $2.80$& $ 5.39$ &
					$0.2250$ & $0.0400$ & $0.00353$ &$2.8\hskip -1 mm\times\hskip -1 mm 10^{-5}$&$66.2^\circ$\\
					$1\,\sigma$	&$\pm 0.10$ &$\pm 1.02$ & $\pm 0.12$& $\pm 1.68$ &$ \pm 0.0007$ &
					$ \pm 0.0008$ & $ \pm 0.00013$ &$^{+0.14}_{-0.12}\hskip -1 mm\times \hskip -1 mm 10^{-5}$&$^{+ 3.4^\circ}_{-3.6^\circ}$\\ \hline  
				\end{tabular}
			\end{center}
		}
			\caption{Results of the fit of the quark mass ratios, 
				CKM mixing angles,  $\delta_{\rm CP}$ and  $J_{\rm CP}$ with complex $g_d$ and $f_u$. 'Exp' denotes the values of the observables at the GUT scale, 
				including $1\sigma$ errors, quoted in Eqs.\,\eqref{Datamass}, 
				\eqref{DataCKM-GUT}, \eqref{CKMphase} and \eqref{JCPGUT}  
				and obtained from the measured ones. 
		}
\label{tab:output-alt4}
\end{table}
%

As it follows from Tables \ref{tab:input-alt4} and \ref{tab:output-alt4}
all quark observables can be successfully reproduced 
with all constant being in magnitude of the same order.
The magnitude of the measure of goodness of the fitting $N\sigma$
is $N\sigma=1.58$.

%
\section{Summary}
\label{sec:Summary}
%

%
We have studied the quark mass hierarchies 
as well as the CKM quark mixing and  CP violation  without 
fine-tuning in a quark flavour model with modular $A_4$ symmetry. 
The quark mass hierarchies are considered close to the 
fixed point $\tau = i\infty$, $\tau$ being the VEV 
of the modulus. In the considered $A_4$ model 
the three left-handed  quark doublets  $Q=((u,d),(c,s),(t,b))$ are assumed to 
furnish a triplet irreducible representation of $A_4$ 
and to carry weight 2, while the three right-handed  down-type quark and 
up-type quark singlets are supposed to be the $A_4$ singlets 
$({\bf 1'},{\bf 1'},{\bf 1'})$ carrying weights (4,\,2,\,0), respectively.
The down-type and up-type quark mass matrices in the model, 
$ M_d$ and $ M_u$, involve modular forms of level 3 and weights 
6, 4 and 2, and each contains four constants, only two ratios of which, $g_d$ 
and $g_u$, can be a source of the CP violation in addition to the VEV of the 
modulus, $\tau$. If  $ M_d$ and $ M_u$ depend on the same $\tau$,
it is possible to reproduce the down-type and up-type 
quark mass hierarchies in the considered model  
for $|g_u|\sim {\cal O}(10)$ with all other constants 
being in magnitude of the same order.
However, reproducing the CP violation in the quark 
sector is problematic. This is due to the fact that the values of 
$|V_{ub}|$ and/or $|V_{cb}|$ elements of the CKM matrix 
resulting from the fit are larger than 
the values of these observables.
We have shown that 
a correct description of the quark mass 
hierarchies, the quark mixing  
and  CP violation is possible close to $\tau = i\infty$ 
with all constant being in magnitude of the same order and complex 
$g_d$ and $g_u$, if there are two different moduli $\tau_d$ and $\tau_u$ 
in the down-type and up-type quark sectors.

We have considered also the case when $ M_d$ and $ M_u$ 
depend on the same $\tau$ and  
involving modular forms of weights 6, 4, 2 and 8, 4, 2, respectively, 
with $ M_u$ receiving a tiny SUSY breaking or higher dimensional operator 
contribution as well.  Both the mass hierarchies of down-type and 
up-type quarks as well and the CKM mixing angles and the CP violating phase 
are  reproduced successfully with one (or two) complex parameter(s) and 
all other parameters being in magnitude of the same order. 
The relatively large value of  ${\rm Im}\,\tau$, needed 
for describing the down-quark mass hierarchies,  
is crucial for obtaining the correct 
up-type quark mass hierarchies.

{ We would like to note that if all theoretical uncertainties
	(due to the scale of SUSY breaking, threshold corrections etc.),
	which are not included in our analyses,
	 are taken into account, the numerical conclusions  may  change.
	Then, it is not excluded that  the models we have found to  fail 
  to describe correctly the data on certain observable 
($|V_{cb}|$, for example) might  actually be successful in 
describing the corresponding observable.
}

{ Finally we comment on the number of parameters of our models.
 	The minimum number of parameters
 	is 12 (10 real constants and 2 phases)  as shown in subsection 6.3.1.
 	Although it is larger than the 10 quark observables, 
the  model is valuable in revealing the possible origin of  
the down-type and up-type quark mass hierarchies.}

The results of our study show that describing correctly 
without severe fine-tuning the quark mass hierarchies, 
the quark mixing and  CP violation in the quark sector 
is remarkably challenging within the modular invariance approach 
to the quark flavour problem.

\section*{Acknowledgments}
The work of S. T. P. was supported in part by the European
Union's Horizon 2020 research and innovation programme under the 
Marie Sk\l{}odowska-Curie grant agreement No.~860881-HIDDeN, by the Italian 
INFN program on Theoretical Astroparticle Physics and by the World Premier 
International Research Center Initiative (WPI Initiative, MEXT), Japan.
The authors would like to thank Kavli IPMU, University of Tokyo, 
where part of this study was performed, for the kind hospitality.

%
\appendix
\section*{Appendix}
%

%
\section{Tensor product of  $\rm A_4$ group}
\label{Tensor}
%

We take the generators of $A_4$ group for the triplet as follows:
\begin{align}
\begin{aligned}
S=\frac{1}{3}
\begin{pmatrix}
-1 & 2 & 2 \\
2 &-1 & 2 \\
2 & 2 &-1
\end{pmatrix},
\end{aligned}
\qquad 
\begin{aligned}
T=
\begin{pmatrix}
1 & 0& 0 \\
0 &\omega& 0 \\
0 & 0 & \omega^2
\end{pmatrix}, 
\end{aligned}
\label{ST}
\end{align}
%
where $\omega=e^{i\frac{2}{3}\pi}$.
In this basis, the multiplication rules are:
\begin{align}
& \begin{pmatrix}
a_1\\
a_2\\
a_3
\end{pmatrix}_{\bf 3}
\otimes 
\begin{pmatrix}
b_1\\
b_2\\
b_3
\end{pmatrix}_{\bf 3} 
\nonumber \\
&=\left (a_1b_1+a_2b_3+a_3b_2\right )_{\bf 1} 
 \oplus \left (a_3b_3+a_1b_2+a_2b_1\right )_{{\bf 1}'}
 \oplus \left (a_2b_2+a_1b_3+a_3b_1\right )_{{\bf 1}''} 
\nonumber \\
&\oplus \frac13
\begin{pmatrix}
2a_1b_1-a_2b_3-a_3b_2 \\
2a_3b_3-a_1b_2-a_2b_1 \\
2a_2b_2-a_1b_3-a_3b_1
\end{pmatrix}_{{\bf 3}}
\oplus \frac12
\begin{pmatrix}
a_2b_3-a_3b_2 \\
a_1b_2-a_2b_1 \\
a_3b_1-a_1b_3
\end{pmatrix}_{{\bf 3}\  } \ ,\\
& {\bf 1} \otimes {\bf 1} = {\bf 1}\,, \qquad 
{\bf 1'} \otimes {\bf 1'} = {\bf 1''}\,, \qquad
{\bf 1''} \otimes {\bf 1''} = {\bf 1'}\,, \qquad
{\bf 1'} \otimes {\bf 1''} = {\bf 1}\,,
\nonumber \\
& {\bf 1'} \otimes
\begin{pmatrix}
a_1\\
a_2\\
a_3
\end{pmatrix}_{\bf 3}
= 
\begin{pmatrix}
a_3\\
a_1\\
a_2
\end{pmatrix}_{\bf 3}\,, \qquad 
{\bf 1''}\otimes
\begin{pmatrix}
a_1\\
a_2\\
a_3
\end{pmatrix}_{\bf 3}
= 
\begin{pmatrix}
a_2\\
a_3\\
a_1
\end{pmatrix}_{\bf 3}\,, 
\end{align}
%
where
\begin{align}
S({\bf 1')}=1\,,\qquad S({\bf 1''})=1, \qquad T({\bf 1')}=\omega\,,\qquad T({\bf 1''})=\omega^2. 
\label{singlet-charge}
\end{align}
%
Further details can be found in the reviews~
\cite{Ishimori:2010au,Ishimori:2012zz,Kobayashi:2022moq}.

\section{A measure of fit}
\label{fit}

As a measure of goodness of fit, we use the sum of one-dimensional 
$\Delta\chi^2$ for eight   observable quantities
$q_j=(m_d/m_b, \, m_s/m_b,\,m_u/m_t, \, m_c/m_t,\, |V_{us}|,\, |V_{cb}|,\, |V_{ub}|,\, \delta_{CP})$.
By employing the Gaussian approximation, we difine
$N\sigma\equiv \sqrt{\Delta \chi^2}$, where
\begin{align}
\Delta \chi^2=
\sum_j \left ( \frac{q_j-q_{j,{\rm best\,fit}}}{\sigma_j}\right )^2\,.
\end{align}
%
%

\section{Massless quark in the mass matrix with weight $(6,2,0)$ }
\label{Det=0}
Consider the mass matrix  by only replacing in Table \ref{tab:model} the weights  $(6,\,2\,,0)$ of the  right-handed  quarks.
Then, the mass matrix is obtained easily as: 
\begin{align}
M_q =v_a
\begin{pmatrix}
\alpha'_q  & 0  & 0  \\
0 & \beta_q & 0 \\
0& 0 & \gamma_q \\
\end{pmatrix}
\begin{pmatrix}
\tilde Y_3^{(8)} &  \tilde Y_2^{(8)} &  \tilde Y_1^{(8)} \\
Y_3^{(4)} &  Y_2^{(4)} &   Y_1^{(4)} \\
Y^{(2)}_3 & Y^{(2)}_2 &Y^{(2)}_1 \\
\end{pmatrix},
\label{Massmatrix-II}
\end{align}
%
where
\begin{align}
\tilde Y_i^{(8)}= f_{q} Y_i^{(8)} +  Y_i^{'(8)}\hskip -1 mm,  \qquad
f_{q} \equiv \alpha_a/\alpha'_a\,.
\end{align}
After putting  $Y_i^{(4)}$, $Y_i^{(8)}$ and $Y_i^{'(8)}$
in Eqs.\,\eqref{weight4} and \eqref{weight8},
we can estimate the determinant of $M_q$ as:
\begin{align}
&{\rm Det} [M_q]=v_q \alpha'_q \beta_q \gamma_q \times \nonumber\\
&(Y_1+Y_2+Y_3)(Y_2^2+2Y_1 Y_3)(2Y_1 Y_2+Y_3^2)
(Y_1^2+Y_2^2+Y_3^2-Y_1Y_2-Y_2Y_3-Y_1Y_3)=0\,,
\end{align}
because of $(Y_2^2+2Y_1 Y_3)=0$  as seen in Eq.\,\eqref{condition}.
We can also check that the matrix is rank 2.
This implies that the mass matrix predicts $m_q = 0$.

\vskip 1 cm

\end{document}